\documentclass[12pt,a4paper]{article}

\usepackage{amsmath} 

\usepackage{hyperref}  
\usepackage{graphicx}
\usepackage{slashed}

\allowdisplaybreaks[3] 

\newcommand{\csw}{c_\mathrm{SW}}

\newcommand{\intdk}{\int\limits_{-\pi}^\pi\frac{\mathrm{d}^4k}{(2\pi)^4}}
\newcommand{\intd}[1]{\int\limits_{-\pi}^\pi\frac{\mathrm{d}^4 #1 }{(2\pi)^4}}

\newcommand{\mr}{\mathrm}
\newcommand{\ri}{\mathrm{i}}

\title{Calculation of $\csw$ at one-loop order for Brillouin fermions}

\author{Maximilian Ammer$\,^{a}$, Stephan D\"urr$\,^{a,b}$}

\date{}

\begin{document}
\maketitle

\begin{center}
${}^a${\sl Department of Physics, University of Wuppertal, 42119 Wuppertal, Germany}\\
${}^b${\sl J\"ulich Supercomputing Centre, Forschungszentrum J\"ulich, 52425 J\"ulich, Germany}
\end{center}

\vspace{10pt}

\begin{abstract}
The Brillouin action is a Wilson-like lattice fermion action with a 81-point stencil, which was found to ameliorate the Wilson action in many respects.
The Sheikholeslami-Wohlert coefficient $\csw$ of the clover improvement term has a perturbative expansion $\csw=\csw^{(0)}+g_0^2\csw^{(1)}+\mathcal{O}(g_0^4)$.
At tree-level $\csw^{(0)}=r$ holds for Wilson and Brillouin fermions alike.
We present the Feynman rules for the Brillouin action in lattice perturbation theory, and employ them to calculate the one-loop coefficient $\csw^{(1)}$ with plaquette or L\"uscher-Weisz gluons.
Numerically its value is found to be about half that of the Wilson action.
\end{abstract}

\section{Introduction}

The (massless) Wilson Dirac operator consists of a term involving the covariant derivative and the covariant Laplace operator (also known as the ``Wilson term'' with $r=1$ canonically)
\begin{align}
D_\mr{W}(x,y)=\sum\limits_\mu \gamma_\mu \nabla^\mathrm{std}_\mu(x,y)-\frac{r}{2}\Delta^\mathrm{std}(x,y)
\end{align}
where $\nabla^\mathrm{std}_\mu$ is the standard two-point derivative and $\Delta^\mathrm{std}$ the standard 9-point Laplacian.
In the Brillouin Dirac operator these derivatives are replaced by the 54-point isotropic derivative $\nabla^\mathrm{iso}_\mu$ and the 81-point Brillouin Laplacian $\Delta^\mathrm{bri}$ as defined in Ref.~\cite{Durr:2010ch}, hence
\begin{align}
D_\mr{B}(x,y)=\sum\limits_\mu \gamma_\mu \nabla^\mathrm{iso}_\mu(x,y)-\frac{r}{2}\Delta^\mathrm{bri}(x,y)\;.
\end{align}
The Brillouin operator takes the explicit form
%
\begin{align}\label{eq:D_B}
D_\mr{B}(x,y)=-r\frac{\lambda_0}{2}\delta(x,y)
&+\sum\limits_{\mu=\pm1}^{\pm 4}\Big(\rho_1\gamma_\mu-r\frac{\lambda_1}{2}\Big)W_\mu(x)\delta(x+\hat{\mu},y)\nonumber\\
&+\sum\limits_{\substack{\mu,\nu=\pm1\\ |\mu|\neq|\nu|}}^{\pm 4}\Big(\rho_2\gamma_\mu-r\frac{\lambda_2}{4}\Big)W_{\mu\nu}(x)\delta(x+\hat{\mu}+\hat{\nu},y)\nonumber\\
&+\sum\limits_{\substack{\mu,\nu,\rho=\pm1 \\ |\mu|\neq|\nu|\neq|\rho|}}^{\pm 4}\Big(\frac{\rho_3}{2}\gamma_\mu-r\frac{\lambda_3}{12}\Big)W_{\mu\nu\rho}(x)\delta(x+\hat{\mu}+\hat{\nu}+\hat{\rho},y)\nonumber\\
&+\sum\limits_{\substack{\mu,\nu,\rho,\sigma=\pm1 \\ |\mu|\neq|\nu|\neq|\rho|\neq|\sigma|}}^{\pm 4}\Big(\frac{\rho_4}{6}\gamma_\mu-r\frac{\lambda_4}{48}\Big)W_{\mu\nu\rho\sigma}(x)\delta(x+\hat{\mu}+\hat{\nu}+\hat{\rho}+\hat{\sigma},y)
\end{align}
where $|\mu|\neq|\nu|\neq\ldots$ is used in a transitive way, i.e.\ the sums are over indices with pairwise different absolute values.
In this way the four sums in (\ref{eq:D_B}) include all \emph{off-axis} points that are one, two, three or four hops away from $x$ (but at most one unit in each direction), and
\begin{align}
W_\mu(x)&=U_\mu(x)\\
W_{\mu\nu}(x)&=\frac{1}{2}\Big(U_\mu(x)U_\nu(x+\hat{\mu})+\mathrm{perm}\Big)\\
W_{\mu\nu\rho}(x)&=\frac{1}{6}\Big(U_\mu(x)U_\nu(x+\hat{\mu})U_\rho(x+\hat{\mu}+\hat{\nu})+\mathrm{perms}\Big)\\
W_{\mu\nu\rho\sigma}(x)&=\frac{1}{24}\Big(U_\mu(x)U_\nu(x+\hat{\mu})U_\rho(x+\hat{\mu}+\hat{\nu})U_\sigma(x+\hat{\mu}+\hat{\nu}+\hat{\rho})+\mathrm{perms}\Big)
\end{align}
are gauge-covariant averages of the connecting paths.

The contributing points in (\ref{eq:D_B}) are weighted by the coefficients
\begin{align}
(\rho_1,\rho_2,\rho_3,\rho_4)&=\frac{1}{432}(64,16,4,1)\label{eq:rhos_bri}\\
(\lambda_0,\lambda_1,\lambda_2,\lambda_3,\lambda_4)&=\frac{1}{64}(-240,8,4,2,1)\label{eq:lambdas_bri}
\end{align}
in the Brillouin action, whereas with
\begin{align}
(\rho_1,\rho_2,\rho_3,\rho_4)&=(1/2,0,0,0)\\
(\lambda_0,\lambda_1,\lambda_2,\lambda_3,\lambda_4)&=(-8,1,0,0,0)\label{eq:lambdas_wil}
\end{align}
we recover the Wilson
\footnote{Other values for the weights $\rho_{1,...,4}$ and $\lambda_{0,...,4}$ are possible, as long as they satisfy certain conditions outlined in Ref.~\cite{Durr:2010ch}.
One example are the ``hypercube fermions'' introduced in Ref.~\cite{Bietenholz:1996pf}.
In the ancillary Mathematica file the weights (\ref{eq:rhos_bri}, \ref{eq:lambdas_bri}), which define the Brillouin fermion, may be overwritten by any desired values.}
action.
The formulation in Eq.~(\ref{eq:D_B}) is chosen for clarity and the purpose of doing perturbative (analytical) calculations.
An effective way to implement the Brillouin operator on modern computer architectures has been presented in Ref.~\cite{Durr:2021iff}.

Like for the Wilson action we can add a clover improvement term to the Brillouin action
\begin{align}\label{eq:S_B_clover}
\mathcal{S}_\mathrm{B}^\text{clover}[\overline{\psi},\psi,U]=
\sum\limits_{x,y}\bar{\psi}(x)D_\mr{B}(x,y)\psi(y)+c_\text{SW}\cdot\sum\limits_x\sum\limits_{\mu<\nu}\bar{\psi}(x)\frac{1}{2}\sigma_{\mu\nu}F_{\mu\nu}(x)\psi(x)\;,
\end{align}
where $\sigma_{\mu\nu}=\frac{\ri}{2}(\gamma_\mu\gamma_\nu-\gamma_\nu\gamma_\mu)$ and $F_{\mu\nu}$ hermitean clover field strength.
If the Sheikholeslami-Wohlert coefficient $\csw$ \cite{Sheikholeslami:1985ij} is tuned correctly, it will remove (in on-shell quantities) all effects at $\mathcal{O}(a)$ coming from the fermion action.
It can be determined non-perturbatively through lattice QCD simulations \cite{Jansen:1995ck,Luscher:1996sc,Luscher:1996ug},
but it turns out that perturbative calculations in the small $g_0$ limit help to fit lattice data and determine $\csw$ as a function of the bare coupling.

Perturbative determinations of $\csw^{(1)}$ for the Wilson fermion action in combination with various gluon actions have been carried out in Refs.~\cite{Wohlert:1987,Luscher:1996vw,Aoki:1998qd,Aoki:2003sj,Horsley:2008ap}.
Our article follows Ref.~\cite{Aoki:2003sj} in methodology, except for a change in the regularization procedure.
This is why we repeat the calculation for Wilson fermions (with both plaquette and L\"uscher-Weisz glue), to demonstrate that we recover the known results.
Our result for Brillouin fermions with plaquette glue and $r=1$ was presented in Ref.~\cite{Ammer:2022qrj}.
We now give results for a range of values of $r\in[0.5,1.5]$, also including the L\"uscher-Weisz gluon action.

The remainder of this article is organized as follows.
In section \ref{sec:feynman_rules} we discuss the Feynman rules needed for the one-loop determination of $\csw$ and how they are derived from the Brillouin action.
In section \ref{sec:self_energy} we use some of these Feynman rules to determine the part of the self energy $\Sigma_0$ which corresponds to the critical mass shift $am_\mathrm{crit}$.
These calculations are carried out for plaquette gluons and L\"uscher-Weisz (i.e.\ tree-level Symanzik-improved) gluons, as the latter are frequently used in numerical simulations.
Section \ref{sec:csw1} finally outlines the calculation of the one-loop value of $\csw$ and presents the results, while Section \ref{sec:summary} gives a summary.

\section{Feynman Rules\label{sec:feynman_rules}}

\begin{figure}[!tb]
\centering
\includegraphics[scale=1.0]{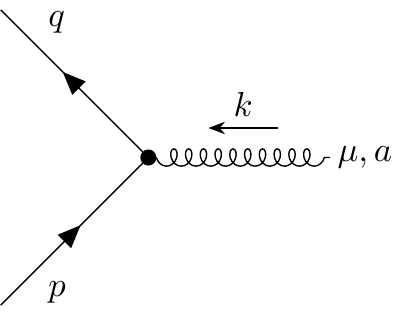}  \hspace{1.0cm}
\includegraphics[scale=1.0]{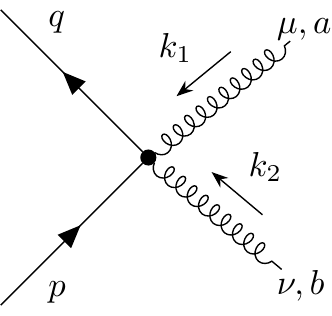} \hspace{1.5cm}
\includegraphics[scale=1.0]{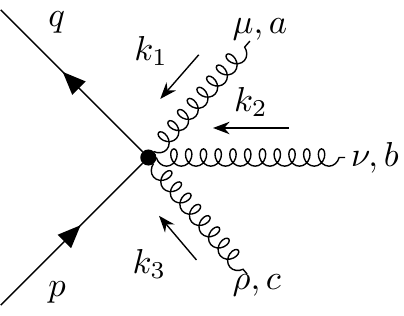}
\caption{Momentum assignments for the vertices with one, two and three gluons.}
\label{fig:vertices}
\end{figure}

The basic idea of lattice perturbation theory is to expand the link variables $U$ in the small coupling limit in terms of the gluon fields $A$
\begin{align}
U_\mu(x)=e^{\ri g_0T^aA^a_\mu(x)}=1+\ri g_0T^aA^a_\mu(x)-\frac{g_0^2}{2}T^aT^bA^a_\mu(x)A^b_\mu(x)+\mathcal{O}(g_0^3)\;.
\end{align}
Inserting this series into the action yields the Feynman rules for vertices of $n$ gluons coupling to a quark anti-quark pair at order $g_0^n$ (contrary to continuum QCD perturbation theory where only the $q\bar{q}g$-vertex exists).
We shall see below that for the one-loop calculation of $\csw$ we need Feynman rules up to order $g_0^3$, which means the $q\bar{q}g$, $q\bar{q}gg$ and $q\bar{q}ggg$-vertices shown in Figure~\ref{fig:vertices}.

After extracting the relevant Feynman rules in position space, we Fourier transform the gluon and fermion fields according to
\begin{align}
A^a_\mu(x)&=\intdk e^{\ri (x+\hat{\mu}/2)k}A^a_\mu(k)\;, &&\\
\psi(x)&=\int\limits_{-\pi}^\pi\frac{\mathrm{d}^4p}{(2\pi)^4} e^{\ri xp}\psi(p)\;,& 
\bar{\psi}(x)&=\int\limits_{-\pi}^\pi\frac{\mathrm{d}^4p}{(2\pi)^4} e^{-\ri xp}\bar{\psi}(p)\\
\delta(x,y)&=\int\limits_{-\pi}^\pi\frac{\mathrm{d}^4p}{(2\pi)^4} e^{\ri (x-y)p}\;,&
\delta(p-q)&=\frac{1}{(2\pi)^4}\sum\limits_x e^{-\ri x(p-q)}
\label{eq:delta_fourier}
\end{align}
where the identities (\ref{eq:delta_fourier}) allow us to perform up to two integrations, thereby enforcing momentum conservation at the vertex (for more detail see Appendix \ref{app:rules}). 
To make the lengthy expressions for the Feynman rules more readable, we use the following shorthand notation for the trigonometric functions
\begin{align}
s(k_\mu)&=\sin(\tfrac 12 k_\mu)& c(k_\mu)&=\cos(\tfrac 12 k_\mu)\\
\bar{s}(k_\mu)&=\sin( k_\mu)& \bar{c}(k_\mu)&=\cos( k_\mu)\\
s^2(k)&=\sum\limits_\mu s(k_\mu)^2& \bar{s}^2(k)&=\sum\limits_\mu \bar{s}(k_\mu)^2\;.
\end{align}

The $q\bar{q}gg$ and $q\bar{q}ggg$-vertices coming from the unimproved ($\csw=0$) Brillouin action are given in Appendix \ref{app:rules}, in Eqs.~(\ref{eq:V2}) and (\ref{eq:V3}), respectvely.
They use functions which are combinations of sine and cosine functions, e.g.\ $K_{\mu\nu}^{(sc)}(p,q)=s(p_\mu+q_\mu)[\bar{c}(p_\nu)+\bar{c}(q_\nu)]$.
The expression for the $q\bar{q}g$-vertex from this part of the action takes the compact form
\begin{align}
V_{1\,\mu}^a(p,q)=-g_0 T^a&\bigg[
r\lambda_1s(p_\mu+q_\mu)+ 2\ri\rho_1 c(p_\mu+q_\mu)\gamma_\mu\nonumber\\
+\sum\limits_{\substack{\nu=1 \\ \nu\neq\mu}}^4 &\Big\{
r\lambda_2 K^{(sc)}_{\mu\nu}(p,q)      
+2\ri\rho_2\big(K^{(cc)}_{\mu\nu}(p,q)\gamma_\mu-K^{(ss)}_{\mu\nu}(p,q)\gamma_\nu\big)\Big\}\nonumber\\
+\frac{1}{3}\sum\limits_{\substack{\nu,\rho=1 \\ \neq(\nu,\rho;\mu)}}^4&  \Big\{
 r\lambda_3 K^{(scc)}_{\mu\nu\rho}(p,q) 
 +2\ri\rho_3 \big(K^{(ccc)}_{\mu\nu\rho}(p,q)\gamma_\mu-2K^{(ssc)}_{\mu\nu\rho}(p,q)\gamma_\nu
\big)\Big\}\nonumber\\
+\frac{1}{9}\sum\limits_{\substack{\nu,\rho,\sigma=1 \\ \neq(\nu,\rho,\sigma;\mu)}}^4 &  \Big\{
r \lambda_4 K^{(sccc)}_{\mu\nu\rho\sigma}(p,q)
+2\ri\rho_4 \big(K^{(cccc)}_{\mu\nu\rho\sigma}(p,q)\gamma_\mu-3K^{(sscc)}_{\mu\nu\rho\sigma}(p,q)\gamma_\nu\big)\Big\}
\bigg]
\label{eq:V1}
\end{align}
with the momentum assignments shown in Figure~\ref{fig:vertices}.
In addition, each of the three vertices has a contribution coming from the clover term which is linear in $\csw^{(0)}$.
For the $q\bar{q}g$-vertex it is
\begin{align}
V_{1c\,\mu}^a(p,q)&=\ri g_0T^a\frac{1}{2}\csw^{(0)}\sum\limits_\nu\sigma_{\mu\nu}c(p_\mu-q_\mu)\bar{s}(p_\nu-q_\nu).
\label{clover_vertex_1}
\end{align}
and the analogous contributions to the $q\bar{q}gg$- and $q\bar{q}ggg$-vertices are given in Appendix \ref{app:clover_rules}. \\
The vertices given here are  not symmetrised with respect to the incoming gluons. We do this at the level of the diagrams, where we sum over all possible connections of internal gluon lines.
%

Furthermore, we need the propagator of the free Brillouin fermion
\begin{align}
&S_\mathrm{B}(k)=\bigg(\sum\limits_\mu \gamma_\mu \nabla_\mu^\mathrm{iso}(k)-\frac{r}{2} \Delta^\mathrm{bri}(k)\bigg)^{-1}
=\frac{-\sum_\mu \gamma_\mu \nabla_\mu^\mathrm{iso}(k)-\frac{r}{2} \Delta^\mathrm{bri}(k)}{\frac{r^2}{4} \Delta^\mathrm{bri}(k)^2-\Big(\sum_\mu  \nabla_\mu^\mathrm{iso}(k)^2\Big)}
\end{align}
with the Fourier transforms of the ``free'' derivative $\nabla_\mu^\mathrm{iso}$ and Laplace operator $\Delta^\mathrm{bri}$
\begin{align}
\nabla_\mu^\mathrm{iso}(k)&=\frac{\ri}{27} \bar{s}(k_\mu)\prod\limits_{\nu\neq \mu}(\bar{c}(k_\nu)+2)\\
\Delta^\mathrm{bri}(k)&=4\Big(c(k_1)^2c(k_2)^2c(k_3)^2c(k_4)^2-1\Big)
\end{align}
whereupon the fermion propagator takes its final form
\begin{align}
&S_\mathrm{B}(k)=
\frac{-\frac{\ri}{27}\sum_\mu\big(\gamma_\mu \bar{s}(k_\mu)\prod_{\nu\neq\mu}(\bar{c}(k_\nu)+2)\big)-2r\big(c(k_1)^2c(k_2)^2c(k_3)^2c(k_4)^2-1\big)}
{4r^2\big(c(k_1)^2c(k_2)^2c(k_3)^2c(k_4)^2-1\big)^2+\frac{1}{729}\sum_\mu\big(\bar{s}(k_\mu)^2\prod_{\nu\neq \mu}(\bar{c}(k_\nu)+2)^2\big)}
\;.
\end{align}

\begin{figure}[!tb]
\centering
\includegraphics[scale=1.0]{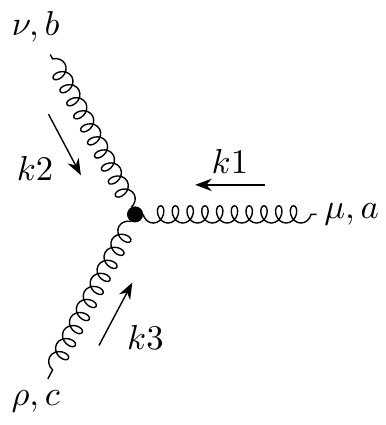}
\caption{The three-gluon vertex.}
\label{fig:Vg3}
\end{figure}

For the gluonic part of the action we consider the standard plaquette action ($c_0=1$, $c_1=0$) as  well as the tree-level Symanzik improved ``L\"uscher-Weisz'' action ($c_0=\frac{5}{3}$, $c_1=-\frac{1}{12}$) \cite{Luscher:1984xn}
\begin{align}
\mathcal{S}_g[U]=-\frac{2}{g_0^2}\sum\limits_{x}\mathrm{Re}\bigg[c_0\sum\limits_\mathrm{plaq}\mathrm{Tr}\big(\mathbf{1}-U^\mathrm{plaq}(x)\big)+
c_1\sum\limits_\mathrm{rect}\mathrm{Tr}\Big(\mathbf{1}-U^\mathrm{rect}(x)\Big)\bigg]
\end{align}
where ``plaq'' stands for all $1\times 1$ plaquettes, ``rect'' for all $1\times 2$ rectangular loops and the coefficients fulfil $c_0+8c_1=1$.
The gluon propagator $G_{\mu\nu}(k)$ and the three-gluon vertex $V_{g3\,\mu\nu\rho}^{abc}(k_1,k_2,k_3)$ for these actions in general covariant gauge are given in Appendix \ref{app:gauge_rules} and were derived in Refs.~\cite{Weisz:1982zw,Weisz:1983bn}.
We follow the convention of inflowing momenta, see Figure~\ref{fig:Vg3}, and use the Feynman gauge ($\alpha=1$) in our calculations. 

We have used a computer algebra system (Mathematica \cite{Mathematica}) to derive the Feynman rules and have additionally checked some of the calculations by hand.

\section{Self-Energy Calculation \label{sec:self_energy}}

\begin{figure}[!tb]
\centering
\includegraphics[scale=1.0]{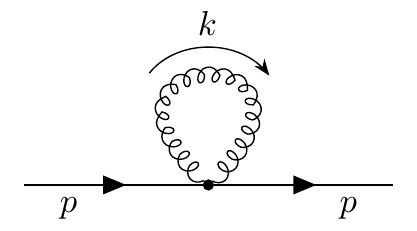}
\includegraphics[scale=1.0]{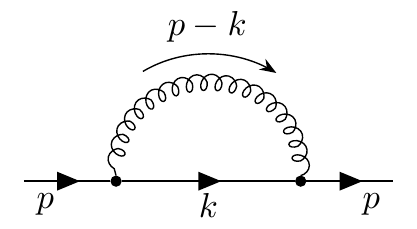}
\caption{Tadpole (left) and sunset (right) diagrams of the quark self energy.}
\label{fig:selfenergy}
\end{figure}

Let us apply the Feynman rules to the calculation of the self-energy of the fermion
\begin{align}
\Sigma=\frac{g_0^2C_F}{16\pi^2}\bigg(\frac{\Sigma_0}{a}+\mathcal{O}(a^0)\bigg)
\end{align}
to first order.
It is given by the ``tadpole'' and ``sunset'' diagrams shown in Figure~\ref{fig:selfenergy}, and yields the additive mass shift
\begin{align}
am_{\mathrm{crit}}=-\frac{g_0^2C_F}{16\pi^2}\Sigma_0
\;.
\end{align}
The contributing integrals for the two diagrams are
\begin{align}
g_0^2C_F\Sigma_0^{\mathrm{(tadpole)}}&=\intdk \sum\limits_{\mu,\nu,a}\Big[ G_{\mu\nu}(k)V_{2\,\mu\nu}^{aa}(p,p,k,-k)\Big]_{p=0}\\
g_0^2C_F\Sigma_0^{\mathrm{(sunset)}}&=\intdk\sum\limits_{\mu,\nu,a}\Big[ V_{1\,\mu}^a(p,k)G_{\mu\nu}(p-k)S(k)V_{1\,\nu}^a(k,p)\Big]_{p=0}
\;.
\end{align}
These are simple and well known integrals in the Wilson/plaquette case (see for example Ref.~\cite{Capitani:2002mp}).
Results for the Wilson/Symanzik case can be found in Ref.~\cite{Horsley:2008ap}.

\begin{table}[!tb]
\centering
\begin{tabular}{|c|c|c|c|}
\hline
Action   & $\Sigma_0^{\mathrm{(tadpole)}}$  & $\Sigma_{0}^{\mathrm{(sunset)}}$& $\Sigma_0$ \\
\hline
Wilson/Plaq.    & $-48.9322(1)$ & $16.9458(1)$ & $-31.9864(1)$ \\
Brillouin/Plaq. & $-48.9322(1)$ & $17.7727(1)$ & $-31.1595(1)$ \\
Wilson/Sym.     & $-40.5177(1)$ & $16.6854(1)$ & $-23.8323(1)$ \\
Brillouin/Sym.  & $-39.0998(1)$ & $16.3742(1)$ & $-22.7256(1)$ \\
\hline
\end{tabular}
\caption{Contributions to the self energy $\Sigma_0$ for $r=1$ and $N_c=3$ coming from the tadpole and sunset diagrams, along with the sum.}
\label{tab:selfenergy}
\end{table}

\begin{table}[!tb]
\centering
\begin{tabular}{|c|c|c|c|}
\hline
Action  & $\Sigma_{00}^{\mathrm{(sunset)}}$ & $\Sigma_{01}^{\mathrm{(sunset)}}$ & $\Sigma_{02}^{\mathrm{(sunset)}}$  \\
\hline
Wilson/Plaq.    & $-2.5025(1)$ & $13.7331(1)$ & $5.7151(1)$ \\
Brillouin/Plaq. & $-5.0086(1)$ & $12.9489(1)$ & $9.8325(1)$ \\
Wilson/Sym.     & $ 0.0745(1)$ & $11.9482(1)$ & $4.6627(1)$ \\
Brillouin/Sym.  & $-2.8534(1)$ & $11.3450(1)$ & $7.8826(1)$ \\
\hline
\end{tabular}
\caption{Contributions to the self energy $\Sigma_0^\mr{(sunset)}$ of the sunset diagram at different orders of the tree level value $\csw^{(0)}$ (for $r=1$ and $N_c=3$).}
\label{tab:sesunset}
\end{table}

Complete results for Wilson or Brillouin fermions with a clover term and plaquette or Symanzik glue are given in Table~\ref{tab:selfenergy}.
The tadpole contribution is exactly the same for Wilson and Brillouin fermions with plaquette glue but not any more identical with the Symanzik-improved L\"uscher-Weisz gluon propagator.
It also gains no additional terms from the inclusion of the clover term.
The sunset contribution, however, does.
It takes the form
\begin{align}
\Sigma_0^\mathrm{(sunset)}=\Sigma_{00}^\mathrm{(sunset)}+\csw^{(0)}\Sigma_{01}^\mathrm{(sunset)}+(\csw^{(0)})^2\,\Sigma_{02}^\mathrm{(sunset)}
\end{align}
where the tree-level value of the improvement coefficient, $\csw^{(0)}=r$, is to be used for a complete one-loop result.
The coefficients $\Sigma_{00}^\mr{(sunset)},\Sigma_{01}^\mr{(sunset)},\Sigma_{02}^\mr{(sunset)}$ are given in Table~\ref{tab:sesunset}.

\begin{figure}[!tb]
\centering
\includegraphics[scale=0.8]{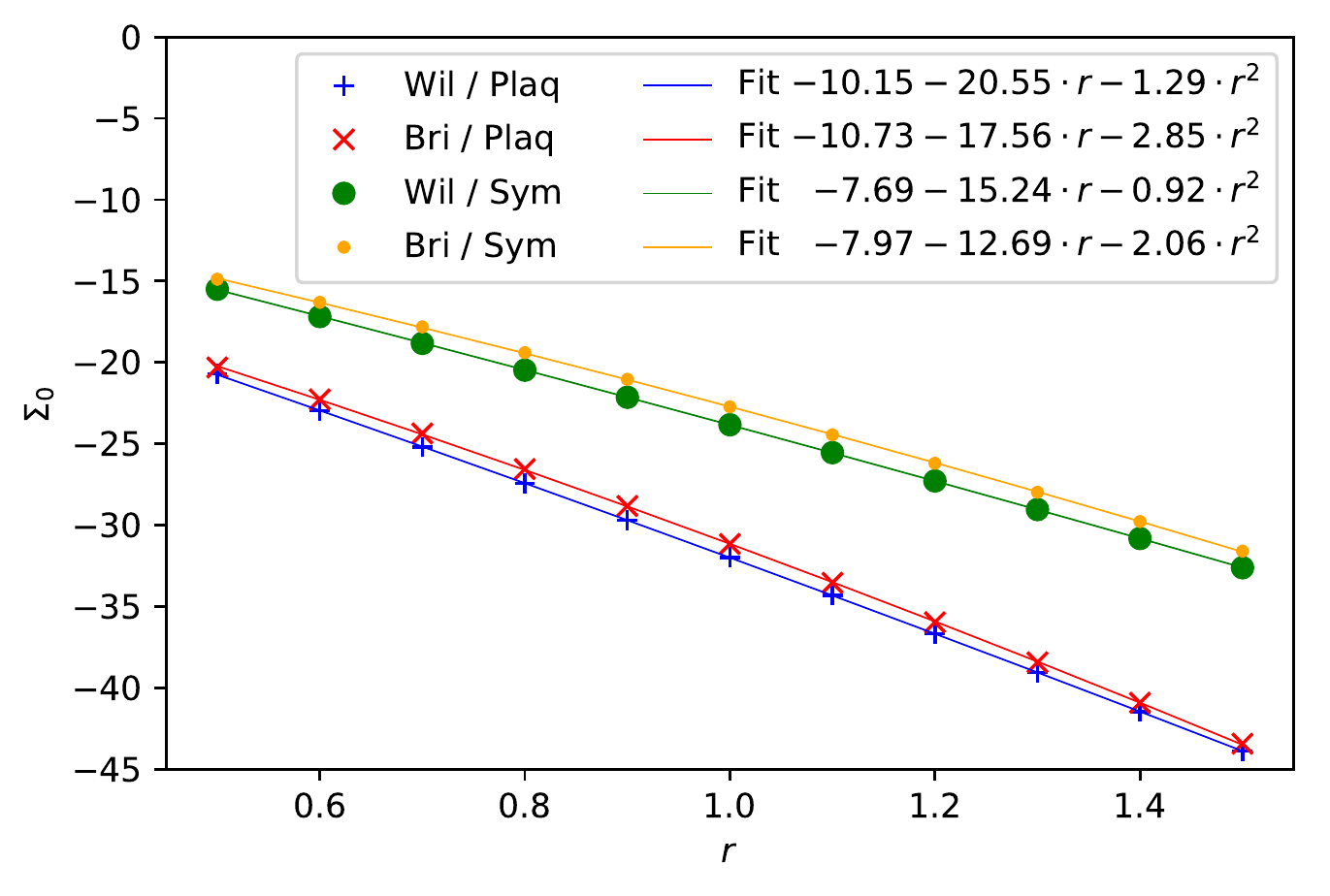}
\vspace*{-8pt}
\caption{Self energy $\Sigma_0$ of Wilson and Brillouin fermions as a function of $r$.}
\label{fig:sigma0_vs_r}
\end{figure}

Figure~\ref{fig:sigma0_vs_r} shows $\Sigma_0$ for values of $r$ ranging from $0.5$ to $1.5$, along with quadratic fit functions for the reader's convenience.
Further details are given in Appendix~\ref{app:self_energy}.
Overall we find that switching from Wilson to Brillouin fermions decreases $|\Sigma_0|\propto am_\mr{crit}$ only marginally, while replacing plaquette gluons by L\"uscher-Weisz gluons has a much more significant effect.

\section{Perturbative Determination of $\csw$ \label{sec:csw1}}

The improvement coefficient $\csw$ in Eq.~(\ref{eq:S_B_clover}) has a perturbative expansion  
\begin{align}
\csw=\csw^{(0)}+g_0^2\csw^{(1)}+\mathcal{O}(g_0^4)
\end{align}
in powers of the bare coupling $g_0^2=2N_c/\beta$. 
It can be calculated via the quark-quark-gluon-vertex function
\begin{align}
\Lambda^a_\mu(p,q)=\sum\limits_{L=0}^\infty g_0^{2L+1} \Lambda^{a(L)}_\mu(p,q)
\label{def_Lambda}
\end{align}
where $L$ is the number of loops.

\subsection{Tree Level}
At the tree level the expression (\ref{def_Lambda}) is given by the lattice $q\bar{q}g$-vertex which is the sum of (\ref{eq:V1}) and \ref{clover_vertex_1}. 
Expanding it to first order in $a$ gives
\begin{align}
&\Lambda^{a(0)}_\mu(p,q)=V_{1\,\mu}^a(ap,aq)
 =-g_0 T^a\bigg(\ri\gamma_\mu\big(2\rho_1+12\rho_2+24\rho_3+16\rho_4\big)   \nonumber\\
&+a \bigg[\frac{r}{2}(p_\mu+q_\mu)\big(\lambda_1+6\lambda_2+12\lambda_3+8\lambda_4\big)+\frac{\ri}{2}\csw \sum\limits_\nu\sigma_{\mu\nu}(p_\nu-q_\nu)\bigg]+\mathcal{O}(a^2)\bigg)
\;,
\label{eq:Lambda0_off_shell}
\end{align}
where $\csw=\csw^{(0)}$ at leading order in $g_0$, and
sandwiching this expression with on-shell spinors $u$ and $\bar{u}$ yields
\begin{align}
\bar{u}(q)\Lambda^{a(0)}_\mu(p,q)u(p)=&-g_0T^a \bar{u}(q)\bigg(\ri\gamma_\mu  \big(2\rho_1+12\rho_2+24\rho_3+16\rho_4\big) \nonumber\\
&+\frac{a}{2}\Big[r\big(\lambda_1+6\lambda_2+12\lambda_3+8\lambda_4\big)-\csw^{(0)}\Big](p_\mu+q_\mu)\bigg)u(p)+\mathcal{O}(a^2)
\end{align}
and hence the condition  $\csw^{(0)}=r\big(\lambda_1+6\lambda_2+12\lambda_3+8\lambda_4\big)$ to eliminate $\mathcal{O}(a)$ contributions at tree level.
From Eqs.~(\ref{eq:lambdas_bri}) and (\ref{eq:lambdas_wil}) we see that both actions fulfill $\big(\lambda_1+6\lambda_2+12\lambda_3+8\lambda_4\big)=1$.
Thus the tree-level improvement coefficient is equal%
\footnote{Some authors write $r\csw$ instead of $\csw$ in Eq.~(\ref{eq:S_B_clover}), such that $\csw^{(0)}$ always equals one.
However, the one-loop value $\csw^{(1)}$ does not naturally factorise in this way, and has a complicated, non-polynomial $r$-dependence.}
to the Wilson parameter $r$ for both Wilson and Brillouin fermions.

\subsection{One-Loop Level: Setup}

At the one-loop level the general form of the vertex function is
\footnote{The function $F_1(p,q)$ gives the continuum result, while $F_2\dots,H_1$ are the limits of Lorentz-invariant functions of the outer momenta $p,q$ for $p,q\to 0$. }
\begin{align}
g_0^3\Lambda^{a(1)}_\mu(p,q) &= -g_0^3 T^{a} \Big(\gamma_\mu F_1(p,q)
 + a \Big[ \slashed{q} \gamma_\mu F_2 + a \gamma_\mu\slashed{p} F_3 +a(p_\mu+q_\mu)G_1  +a(p_\mu-q_\mu)H_1 \nonumber\\
&+\mathcal{O}(p^2,q^2,pq)\Big]+\mathcal{O}(a^2) \Big)
\label{eq:expansion_of_Lambda}
\end{align}
where the $F_2$, $F_3$ and $H_1$ terms do not contribute to on-shell quantities \cite{Aoki:2003sj}.

At order $g_0^3$ Eq.~(\ref{eq:Lambda0_off_shell}) with $\csw=\csw^{(0)}+g_0^2\csw^{(1)}+\hdots$ contributes a single term, involving $\csw^{(1)}$. Together with the one-loop vertex function (\ref{eq:expansion_of_Lambda}) and going on-shell we get 
\begin{align}
&g_0^3\bar{u}(q)\Big(\frac{a}{2}\csw^{(1)}(p_\mu+q_\mu)T^a + \Lambda^{a(1)}_\mu(p,q) \Big)u(p)\nonumber\\
=&g_0^3\bar{u}(q)T^a\Big(-\gamma_\mu F_1+ \frac{a}{2}(p_\mu+q_\mu)(\csw^{(1)}-2G_1)  +\mathcal{O}(a^2) \Big)u(p)
\end{align}
and this results in the condition
\begin{align}
\csw^{(1)}=2G_1
\end{align}
to eliminate $\mathcal{O}(a)$ contributions at the one-loop level.
We use the following equation (given in Ref.~\cite{Aoki:2003sj}) to extract $G_1$ from the off-shell vertex function
\begin{align}\label{eq:G1}
g_0^3T^aG_1=-\frac{1}{8}\text{Tr}
\bigg[\bigg(\frac{\partial}{\partial p_\mu}+\frac{\partial}{\partial q_\mu}\bigg)\Lambda^{a(1)}_\mu - \bigg(\frac{\partial}{\partial p_\nu}-\frac{\partial}{\partial q_\nu}\bigg)\Lambda^{a(1)}_\mu\gamma_\nu\gamma_\mu  \bigg]^{\mu\neq\nu}_{p,q\rightarrow 0}
\;.
\end{align}

\begin{figure}[!tb]
\centering
\begin{tabular}{ccc}
 (a) & (b) & (c) \\
\includegraphics[scale=1.0]{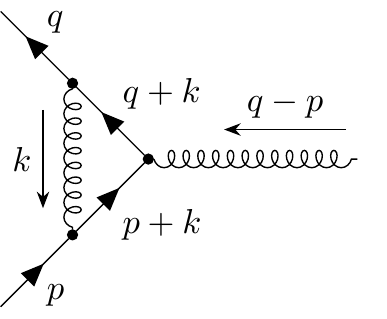} &
\includegraphics[scale=1.0]{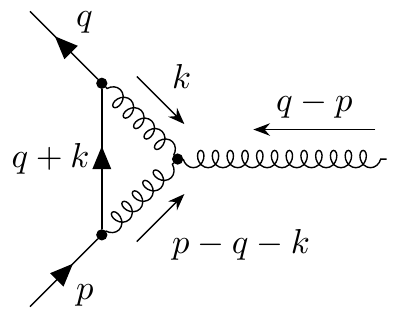} &
\includegraphics[scale=1.0]{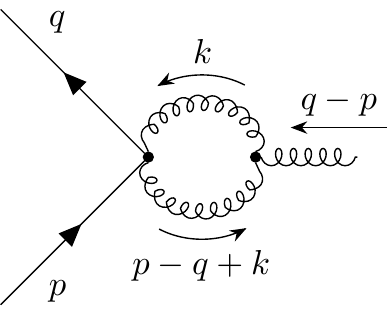} \\
 (d) & (e) & (f) \\
\includegraphics[scale=1.0]{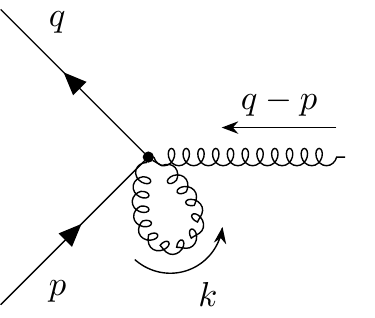} &
\includegraphics[scale=1.0]{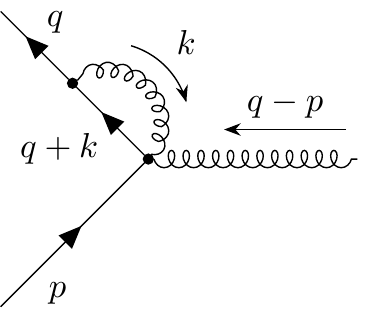} &
\includegraphics[scale=1.0]{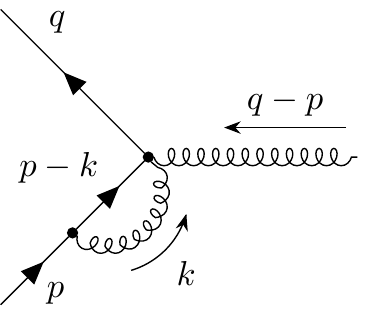}
\end{tabular}
\caption{The six one-loop diagrams contributing to the vertex function.}
\label{fig:diagrams}
\end{figure}

The six diagrams that contribute at the one-loop level are shown in Figure~\ref{fig:diagrams}.

We use Mathematica notebooks that take the previously generated Feynman rules as input and construct from them the relevant integrals corresponding to the six diagrams.
Next, Eq.~(\ref{eq:G1}) gets applied to each integral, and the resulting expressions are added and numerically integrated over the four-dimensional Brillouin zone.
For individual diagrams the integrals (with one exception) are divergent and need to be regularised, which is explained below.

\subsection{One-loop level: Regularisation}

The sum of all diagrams is finite and can be numerically integrated without further ado (``all in one approach'').
As a check and to compare results with other calculations it is very helpful to calculate the contribution from each diagram separately.
Five of the six diagrams are logarithmically IR-divergent.
Therefore we need to regularise these integrals and extract the divergent part before we can evaluate them numerically.
To this end we subtract from each divergent lattice integral the simplest logarithmically divergent lattice integral
\begin{align}
\mathcal{B}_2=\intdk \frac{1}{(4s^2(k))^2}
\label{def_mathcalb2}
\end{align}
multiplied with an appropriate pre-factor.
The integral $\mathcal{B}_2$ is regularised by a small fictitious gluon mass $\mu$
\begin{align}
\mathcal{B}_2(\mu)=\frac{1}{16\pi^2}\big(-\ln(\mu^2)+F_0-\gamma_E\big)+\mathcal{O}(\mu^2)
\label{val_mathcalb2}
\end{align}
where $\gamma_E=0.57721566490153(1)$ is the Euler-Mascheroni constant and $F_0$ a lattice constant given in Ref.~\cite{Burgio:1996ji} as $F_0=4.369225233874758(1)$.

Because in the sum of all diagrams the divergencies cancel, we can add arbitrary constants to $\mathcal{B}_2$.
We like to compare our results to those by Aoki and Kuramashi in Ref.~\cite{Aoki:2003sj}, who calculated $\csw^{(1)}$ for the Wilson action and for various improved gauge actions.
They used a slightly different method of regularisation and encoded the divergence in
\begin{align}
L=\frac{1}{16\pi^2}\ln\bigg(\frac{\pi^2}{\mu^2}\bigg)
\;.
\end{align}
In essence they used the following two divergent \emph{continuum}-like integrals
\begin{align}
\mathcal{L}_1&=\intdk\frac{1}{k^2(k^2+\mu^2)} =L+\mathcal{O}(\mu^2) \\
\mathcal{L}_2&=\intdk\frac{1}{(k^2+\mu^2)^2}=L-\frac{1}{16\pi^2}+\mathcal{O}(\mu^2)
\end{align}
coming from the expansion of the lattice Feynman rules to $\mathcal{O}(a)$.
The integral $\mathcal{L}_1$ appears in diagrams (a),(e) and (f), whereas $\mathcal{L}_2$ appears in diagrams (b) and (c).
Hence the relations
\begin{align}
\mathcal{L}_1&=\mathcal{B}_2(\mu)+\frac{1}{16\pi^2}\big(\ln(\pi^2)-F_0+\gamma_E\big)
\label{translation1}
\\
\mathcal{L}_2&=\mathcal{B}_2(\mu)+\frac{1}{16\pi^2}\big(\ln(\pi^2)-F_0+\gamma_E-1\big)
\label{translation2}
\end{align}
will convert our results to those in the formalism of Ref.~\cite{Aoki:2003sj} (see Table~\ref{tab:results1} and Table~\ref{tab:results2} below).

With this setup the calculation is straight-forward, though tedious.
As mentioned before, we use Mathematica to carry out the algebraic manipulations \cite{Mathematica} and the files are available as ancillary material.
In case of diagram (c) the calculation can be carried out by hand, the essential steps being given in Appendix~\ref{app:dia_c}.

\subsection{One-loop level: Results}

The numerical results for each diagram and their sum is given in Table \ref{tab:results1} for Wilson and Brillouin fermions, with plaquette and L\"uscher-Weisz glue.
Table \ref{tab:results2} contains the same results converted into the formalism of Ref.~\cite{Aoki:2003sj}.
Our numbers in the ``Wilson'' columns agree with (and improve on) the numbers given there.
Throughout the ``sum'' (which results from adding the six regulated single-diagram contributions) agrees with the result of the direct numerical integration of the unregularized (finite) ``all in one'' approach.

\begin{table}[!tb]
\centering
\def\arraystretch{1.}
\begin{tabular}{|c|c|c|c|c|c|}
\hline
Diag. & $\mathcal{B}_2$ & Wilson/Plaq. & Brillouin/Plaq. & Wilson/Sym. & Brillouin/Sym.       \\
\hline
(a)   & $-1/3$ & $ 0.009852153(1)$ & $ 0.0100402212(1)$ & $ 0.01048401(1)$  & $ 0.0108335(1)$ \\
(b)   & $-9/2$ & $ 0.125895883(1)$ & $ 0.098371668(1)$  & $ 0.1285594(1)$   & $ 0.102829(1)$  \\
(c)   & $+9/2$ & $-0.124125079(1)$ & $-0.100558858(1)$  & $-0.1337781(1)$   & $-0.1098254(1)$ \\
(d)   & $0$    & $ 0.297394534(1)$ & $ 0.142461144(1)$  & $ 0.2354388(1)$   & $ 0.1120815(1)$ \\
(e)   & $+1/6$ & $-0.020214623(1)$ & $-0.013344189(1)$  & $-0.022229808(1)$ & $-0.013659(1)$  \\
(f)   & $+1/6$ & $-0.020214623(1)$ & $-0.013344189(1)$  & $-0.022229808(1)$ & $-0.013659(1)$  \\
\hline
Sum   & $0$    & $ 0.26858825 (1)$ & $ 0.12362580 (1)$  & $ 0.1962445(1)$   & $0.088601(1)$   \\
\hline
\end{tabular}
\caption{Divergent and constant contributions to $\csw^{(1)}$ from each diagram for $N_c=3$ and $r=1$.
The second column gives the coefficients in front of the logarithmically divergent $\mathcal{B}_2(\mu)=\frac{1}{16\pi^2}\big(-\ln(\mu^2)+F_0-\gamma_E\big)$.}
\label{tab:results1}
\end{table}

\begin{table}[!tb]
\centering
\def\arraystretch{1.}
\begin{tabular}{|c|c|c|c|c|c|}
\hline
Diag. & $L$    & Wilson/Plaq. & Brillouin/Plaq. & Wilson/Sym. & Brillouin/Sym. \\
\hline
(a) & $-1/3$   & $ 0.004569626(1)$ & $ 0.0047576939(1)$ & $ 0.00520148(1)$  & $0.0055510(1)$  \\
(b) & $-9/2$   & $ 0.083078349(1)$ & $ 0.055554134(1)$  & $ 0.1142384(1)$   & $0.088508(1)$   \\
(c) & $+9/2$   & $-0.081307544(1)$ & $-0.057741323(1)$  & $-0.1194571(1)$   & $-0.0955044(1)$ \\
(d) & $0$      & $ 0.297394534(1)$ & $ 0.142461144(1)$  & $ 0.2354388(1)$   & $0.1120815(1)$  \\
(e) & $+1/6$   & $-0.017573359(1)$ & $-0.010702925(1)$  & $-0.019588545(1)$ & $-0.011018(1)$  \\
(f) & $+1/6$   & $-0.017573359(1)$ & $-0.010702925(1)$  & $-0.019588545(1)$ & $-0.011018(1)$  \\
\hline
Sum & $0$      & $ 0.26858825(1)$  & $ 0.12362580(1)$   & $0.1962445(1)$    & $0.088601(1)$   \\
\hline
\end{tabular}
\caption{Divergent and constant contributions to $\csw^{(1)}$ from each diagram for $N_c=3$ and $r=1$.
The second column gives the coefficients in front of the logarithmically divergent $L\equiv\frac{1}{16\pi^2}\ln\big(\pi^2/\mu^2\big)$.
This corresponds to the formalism used in Ref.~\cite{Aoki:2003sj}}.
\label{tab:results2}
\end{table}

A closer look at Table~\ref{tab:results1} reveals interesting features.
The second column indicates that not only the divergent parts of all diagrams would cancel each other, but also that that of diagram (a) would cancel the combined (e,f) divergence, and that (b) cancels (c).
This allows us to split the sum of all diagrams into two finite parts, one proportional to $N_c$ [diagrams (a),(d),(e),(f)] and one proportional to $1/N_c$ [diagrams (b),(c),(d)].
The corresponding coefficients are given in Table \ref{tab:fullresult} for $r=1$, and setting $N_c=3$ is found to reproduce the previous results for $\csw^{(1)}$.

\begin{table}[!tb]
\centering
\begin{tabular}{|c|c|c|c|}
\hline
Action           & $N_c$            & $1/N_c$          & $\csw^{(1)}$ for $N_c=3$ \\
\hline
Wilson/Plaq.     & $0.098842471(1)$ & $-0.08381750(1)$ & $0.26858825(1)$          \\
Wilson/Sym.      & $0.0718695(1)$   & $-0.05809245(1)$ & $0.1962445(1)$           \\
Brillouin/Plaq.  & $0.04578552 (1)$ & $-0.04119226(1)$ & $0.12362580(1)$          \\
Brillouin/Sym.   & $0.032600(1)$    & $-0.0275974(1)$  & $0.088601(1)$            \\
\hline
\end{tabular}
\caption{Coefficients of $N_c$ and $1/N_c$ in $\csw^{(1)}$ and the full result for $N_c=3$ for $r=1$.}
\label{tab:fullresult}
\end{table}

\begin{figure}[!tb]
\centering
\includegraphics[scale=0.8]{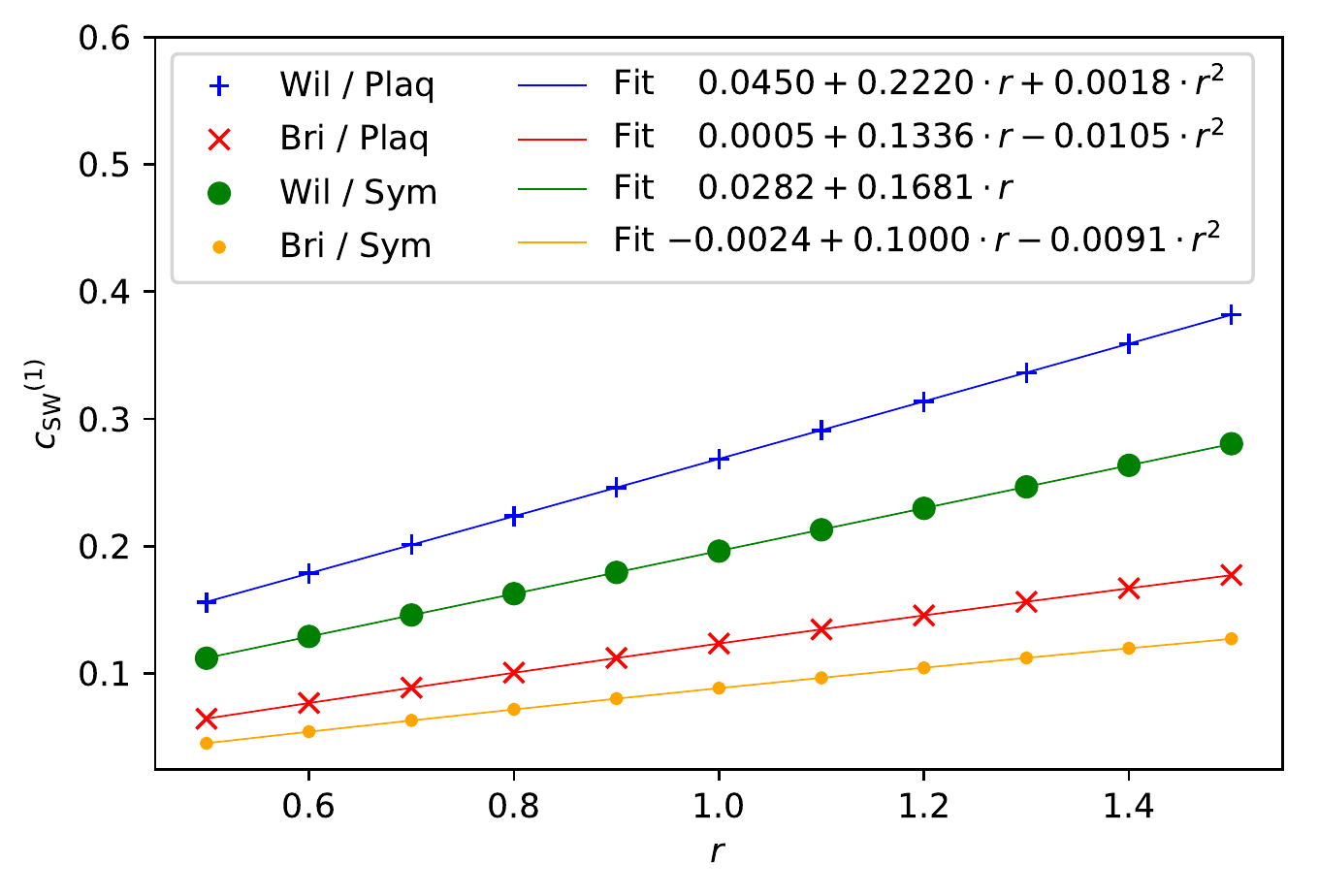}
\caption{The one-loop values of $\csw^{(1)}$ for Wilson and Brillouin fermions with $N_c=3$ as a function of $r$.}
\label{fig:csw1_vs_r}
\end{figure}

For any $N_c$ the value of $\csw^{(1)}$ for Brillouin fermions is about half the value for Wilson fermions.
This holds with plaquette glue and with L\"uscher-Weisz glue.
In addition, we have performed the same calculations for several values of the Wilson parameter $r$ in the range $0.5, 0.6, ... , 1.5$.
The results are shown in Figure~\ref{fig:csw1_vs_r} and listed in the tables of Appendix~\ref{app:csw_r}.

\section{Summary\label{sec:summary}}

This paper provides Feynman rules for the Brillouin fermion action, and lists the vertices $q\bar{q}g$, $q\bar{q}gg$, $q\bar{q}ggg$ that come from the unimproved part of the fermion action.
In addition, each one of these vertices receives a contribution from the clover term which is proportional to $c_\mr{SW}^{(0)}$ and known from Refs.~\cite{Aoki:2003sj,Horsley:2008ap},
since it is the same for Wilson and Brillouin fermions.

As a warm-up exercise we calculate the self-energy contribution $\Sigma_0\propto1/a$ which encodes the critical mass $am_\mr{crit}$ of Brillouin fermions.
We do this for the plaquette gluon action and the L\"uscher-Weisz gluon action, with the Wilson parameter in the range $r\in\{0.5, 0.6, ... , 1.5\}$.
We find that the critical mass nearly matches the one of Wilson fermions, and the tree-level improvement in the gluon sector reduces it (for both fermion actions) by about $25\%$.

Last but not least the well-known tree-level result $c_\mr{SW}^{(0)}=r$ is complemented with the one-loop result $c_\mr{SW}^{(1)}$ for Brillouin fermions.
The sum of the six diagrams is finite and ready for numerical integration.
On a diagram-by-diagram basis, five of the diagrams are IR-divergent and need regularization.
We propose a slightly different regularization technique than Ref.~\cite{Aoki:2003sj}, but the ``translation rules'' Eqs.~(\ref{translation1},~\ref{translation2})
establish complete agreement of our results for the Wilson action with those in the literature, except that it is easier to reach good numerical precision with our quantity 
$\mathcal{B}_2$ as defined in Eqs.~(\ref{def_mathcalb2},~\ref{val_mathcalb2}).

The results are given in Section~\ref{sec:csw1} and Appendix~\ref{app:csw_r} for the range $r\in[0.5,1.5]$ of the Wilson lifting parameter $r$ and arbitrary $N_c$.
These results are ready for use in numerical simulations, and will reduce the leading cut-off effects from $\mathcal{O}(a\log(a))$ to $\mathcal{O}(a\log^2(a))$.
Of course our hope is that these results will eventually be complemented with a non-perturbative determination of $\csw$ for a variety of couplings.
We envisage that future (non-perturbative) data for $c_\mr{SW}$ of the Brillouin action will be fitted with a rational ansatz of the form $c_\mr{SW}=(1+c_1g_0^2+c_2g_0^4+...)/(1+d_1g_0^2+...)$,
where the difference $c_1-d_1$ is restricted to obey our 1-loop perturbative calculation.
This is exactly what was done for Wilson fermions in Refs.~\cite{Jansen:1995ck,Luscher:1996sc,Luscher:1996ug},
and we hope that the improvement program for Brillouin fermions will be similarly successful.
%
\vspace{0.1cm}\\
\textbf{Acknowledgement}\\
We thank Stefano Capitani and Johannes Weber for useful discussion related to this work.
%
\pagebreak
\appendix
\section{Feynman Rules \label{app:rules}}
\subsection{Feynman Rules of the Brillouin Action\label{app:bri_rules}}

\subsubsection{The $\mathbf{q\bar{q}g}$-vertex}

As a short example we show in some detail how to derive the part of the $q\bar{q}g$-vertex $V^a_{1\,\mu}(p,q)$ proportional to $\lambda_2$ and $\rho_2$,
i.e.\ we start with the following term of the Brillouin action
\begin{align}
\sum\limits_{x,y}\bar{\psi}(x)\Bigg(\sum\limits_{\substack{\mu,\nu=\pm1\\ |\mu|\neq|\nu|}}^{\pm 4}
\bigg(\rho_2\gamma_\mu-r\frac{\lambda_2}{4}\bigg)W_{\mu\nu}(x)\delta(x+\hat{\mu}+\hat{\nu},y)\Bigg)\psi(y)
\;.
\end{align}
First we rewrite the sum, such that it is over positive indices only
\begin{align}
&\sum\limits_{x,y}
\sum\limits_{\substack{\mu,\nu=1\\ \mu\neq\nu}}^{4}\bar{\psi}(x)
\Bigg[
\bigg(\rho_2\gamma_\mu-r\frac{\lambda_2}{4}\bigg)
\bigg(W_{\mu\nu}(x)\delta(x+\hat{\mu}+\hat{\nu},y)
+W_{\mu-\nu}(x)\delta(x+\hat{\mu}-\hat{\nu},y)
\bigg)\nonumber\\
&
+\bigg(-\rho_2\gamma_\mu-r\frac{\lambda_2}{4}\bigg)
\bigg(
W_{-\mu\nu}(x)\delta(x-\hat{\mu}+\hat{\nu},y)
+W_{-\mu-\nu}(x)\delta(x-\hat{\mu}-\hat{\nu},y)
\bigg)\Bigg]
\psi(y)
\end{align}
where $\gamma_{-\mu}=-\gamma_\mu$ has been used.
We insert
\begin{align}
W_{\mu\nu}(x)&=\frac{1}{2}\big(U_\mu(x)U_\nu(x+\hat{\mu})+U_\nu(x)U_\mu(x+\hat{\nu})\big)
\end{align}
expanded to order $g_0$, hence we replace
\begin{align}
W_{\mu\nu}(x)  & \to ig_0T^a\frac{1}{2}\big(A^a_\mu(x)+A^a_\nu(x+\hat{\mu})+A^a_\nu(x)+A^a_\mu(x+\hat{\nu})\big)\\
W_{\mu-\nu}(x) & \to ig_0T^a\frac{1}{2}\big(A^a_\mu(x)-A^a_\nu(x+\hat{\mu}-\hat{\nu})-A^a_\nu(x-\hat{\nu})+A^a_\mu(x-\hat{\nu})\big)\\
W_{-\mu\nu}(x) & \to ig_0T^a\frac{1}{2}\big(-A^a_\mu(x-\hat{\mu})+A^a_\nu(x-\hat{\mu})+A^a_\nu(x)-A^a_\mu(x-\hat{\mu}+\hat{\nu})\big)\\
W_{-\mu-\nu}(x)& \to ig_0T^a\frac{1}{2}\big(-A^a_\mu(x-\hat{\mu})-A^a_\nu(x-\hat{\mu}-\hat{\nu})-A^a_\nu(x-\hat{\nu})-A^a_\mu(x-\hat{\mu}-\hat{\nu})\big)
\end{align}
where we have used
\begin{align}
U_{-\mu}(x)=U^\dagger_\mu(x-\hat{\mu})=1-\ri g_0T^aA^a_\mu(x-\hat{\mu})+\mathcal{O}(g_0^2)
\;.
\end{align}
Next we insert the Fourier transforms of $\bar{\psi}$, $\psi$, $A$ and $\delta$.
One example term reads
\begin{align}
&\sum\limits_{x,y}
\sum\limits_{\substack{\mu,\nu=1\\ \mu\neq\nu}}^{4}
\bar{\psi}(x)
\bigg(\rho_2\gamma_\mu-r\frac{\lambda_2}{4}\bigg)
A^a_\mu(x)\delta(x+\hat{\mu}+\hat{\nu},y)\psi(y)\nonumber\\
=&\sum\limits_{x,y}
\sum\limits_{\substack{\mu,\nu=1\\ \mu\neq\nu}}^{4}
\intd{q}\intd{k}\intd{r}\intd{p}\\
&\times
e^{-\ri xq}\bar{\psi}(q)
\bigg(\rho_2\gamma_\mu-r\frac{\lambda_2}{4}\bigg)
e^{\ri (x+\hat{\mu}/2)k} A^a_\mu(k)
e^{\ri (x+\hat{\mu}+\hat{\nu}-y)r}
e^{\ri yp}\psi(p)
\;.
\end{align}
The sum over $y$ introduces $(2\pi)^4\delta(r-p)$, which lets us perform the $r$-integral.
Then the sum over $x$ introduces $(2\pi)^4\delta(q-p-k)$, which enforces momentum conservation at the vertex and allows us to perform the $k$-integral.
As a result we are left with
\begin{align}
\sum\limits_{\substack{\mu,\nu=1\\ \mu\neq\nu}}^{4}
\intd{q}\intd{p}
\bar{\psi}(q)
\bigg(\rho_2\gamma_\mu-r\frac{\lambda_2}{4}\bigg)
\psi(p)A^a_\mu(q-p)
e^{\frac{\ri}{2}(q_\mu+p_\mu)}e^{\ri p_\nu}
\end{align}
and together with the term $-\big(-\rho_2\gamma_\mu-r\frac{\lambda_2}{4}\big)A^a_\mu(x-\hat{\mu})\delta(x-\hat{\mu}-\hat{\nu},y)$ this yields the contribution
\begin{align}
\rho_2\gamma_\mu \cos\big(\tfrac{1}{2}(p_\mu+q_\mu)+p_\nu\big)-\frac{r\lambda_2}{4}\sin\big(\tfrac{1}{2}(p_\mu+q_\mu)+p_\nu\big)
\end{align}
to the $q\bar{q}g$ vertex.
Adding all the other terms and some more trigonometric manipulations will finally result in what is shown in Eq.~(\ref{eq:V1}).

The procedure for the $q\bar{q}gg$- and $q\bar{q}ggg$-vertices is basically the same.
After Fourier transforming the fields, we integrate out the gluon momentum $k_1$ and find ourselves left with expressions containing $p,q,k_2$ and $p,q,k_2,k_3$ respectively.
At this point some diligence is needed to rename the indices in terms containing for example $A^a_\nu(k_1)A^b_\mu(k_2)$ or $A^a_\rho(k_1)A^b_\nu(k_2)A^c_\mu(k_3)$ such that everything is proportional to $A^a_\mu(k_1)A^b_\nu(k_2)$ and $ A^a_\mu(k_1)A^b_\nu(k_2)A^c_\rho(k_3)$, respectively.

\subsubsection{Notation Details}

The following combinations of sine and cosine functions are very convenient to shorten the lengthy expressions for the vertices
\begin{align}
K^{(fg)}_{\mu\nu}(p,q)&=f(p_\mu+q_\mu)\big[\bar{g}(p_\nu)+\bar{g}(q_\nu)\big]\\
K^{(fgh)}_{\mu\nu\rho}(p,q)&=f(p_\mu+q_\mu)\Big\{\bar{g}(p_\nu)\bar{h}(p_\rho)+\bar{g}(q_\nu)\bar{h}(q_\rho)
\nonumber\\
&+\big[\bar{g}(p_\nu)+\bar{g}(q_\nu)\big]\big[\bar{h}(p_\rho)+\bar{h}(q_\rho)\big]\Big\}
\\
K^{(fghj)}_{\mu\nu\rho\sigma}(p,q)&=
f(p_\mu+q_\mu)\Big\{2\big[\bar{g}(p_\nu)\bar{h}(p_\rho)\bar{j}(p_\sigma)+\bar{g}(q_\nu)\bar{h}(q_\rho)\bar{j}(q_\sigma)\big]
\nonumber\\
&+\big[\bar{g}(p_\nu)+\bar{g}(q_\nu)\big]\big[\bar{h}(p_\rho)+\bar{h}(q_\rho)\big]\big[\bar{j}(p_\sigma)+\bar{j}(q_\sigma)\big]\Big\}
\\
L^{(fg)}_{\mu\nu}(p,q,k)&=
f(p_\mu+q_\mu+k_{\mu})g(2p_\nu+k_{\nu})
\\
L^{(fgh)}_{\mu\nu\rho}(p,q,k)&=
f(p_\mu+q_\mu+k_{\mu})g(2p_\nu+k_{\nu})
\big[\bar{h}(p_\rho)+\bar{h}(p_\rho+k_{\rho})+\bar{h}(q_\rho)\big]
\\
L^{(fghj)}_{\mu\nu\rho\sigma}(p,q,k)&=
f(p_\mu+q_\mu+k_{\mu})g(2p_\nu+k_{\nu})
\nonumber\\
&\times\Big\{\bar{h}(p_\rho)\bar{j}(p_\sigma)+\bar{h}(p_\rho+k_\rho)\bar{j}(p_\sigma+k_\sigma)+\bar{h}(q_\rho)\bar{j}(q_\sigma)
\nonumber\\
&+\big[\bar{h}(p_\rho)+\bar{h}(p_\rho+k_\rho)+\bar{h}(q_\rho)\big]\big[\bar{j}(p_\sigma)+\bar{j}(p_\sigma+k_\sigma)+\bar{j}(q_\sigma)\big]\Big\}
\\
M^{(fgh)}_{\mu\nu\rho}(p,q,k_1,k_2)&=
f(p_\mu+q_\mu+k_{1\mu}+k_{2\mu})g(2p_\nu+k_{1\nu}+2k_{2\nu})h(2p_\rho+k_{2\rho})
\\
M^{(fghj)}_{\mu\nu\rho\sigma}(p,q,k_1,k_2)&=
f(p_\mu+q_\mu+k_{1\mu}+k_{2\mu})g(2p_\nu+k_{1\nu}+2k_{2\nu})h(2p_\rho+k_{2\rho})
\nonumber\\
&\times\Big\{\bar{j}(p_\sigma)+\bar{j}(p_\sigma+k_{2\sigma})+\bar{j}(p_\sigma+k_{1\sigma}+k_{2\sigma})+\bar{j}(q_\sigma)\Big\}
\end{align}
with $f,g,h,j\in \{s,c\}$.

\subsubsection{The $\mathbf{q\bar{q}gg}$-vertex}

The $q\bar{q}gg$ vertex for the Brillouin action takes the form
\begin{align}\label{eq:V2}
V_{2\,\mu\nu}^{ab}&(p,q,k_1,k_2)=ag_0^2 T^aT^b\Bigg\{
-\frac{1}{2}r\lambda_1\delta_{\mu\nu}c(p_\mu+q_\mu)+\ri \rho_1 \delta_{\mu\nu}s(p_\mu+q_\mu)\gamma_\mu
\nonumber\\
+r\lambda_2&\bigg((1-\delta_{\mu\nu})L^{(ss)}_{\mu\nu}(p,q,k_2)-
\frac{1}{2}\delta_{\mu\nu}\sum\limits_{\substack{\alpha=1 \\ \alpha\neq\mu}}^4K^{(cc)}_{\mu\alpha}(p,q)\bigg)
\nonumber\\
+r\lambda_3&\bigg(
\frac{2}{3}(1-\delta_{\mu\nu})
\sum\limits_{\substack{\rho=1 \\ \neq(\rho,\mu,\nu)}}^4 
L^{(ssc)}_{\mu\nu\rho}(p,q,k_2)
-\frac{1}{6}\delta_{\mu\nu}\sum\limits_{\substack{\alpha,\rho=1 \\ \neq(\alpha,\rho,\mu)}}^4K^{(ccc)}_{\mu\alpha\rho}(p,q)
\bigg)
\nonumber\\
+r\lambda_4&\bigg(
\frac{1}{6}(1-\delta_{\mu\nu})
\sum\limits_{\substack{\rho,\sigma=1 \\ \neq(\rho,\sigma,\mu,\nu)}}^4 
L^{(sscc)}_{\mu\nu\rho\sigma}(p,q,k_2)
-\frac{1}{18} \delta_{\mu\nu}
\sum\limits_{\substack{\alpha,\rho,\sigma=1 \\ \neq(\alpha,\rho,\sigma,\mu)}}^4
K^{(cccc)}_{\mu\alpha\rho\sigma}(p,q)
\bigg)
\nonumber\\
+\ri \rho_2&\bigg(
2(1-\delta_{\mu\nu})
\Big[L^{(cs)}_{\mu\nu}(p,q,k_2)\gamma_\mu+L^{(sc)}_{\mu\nu}(p,q,k_2)\gamma_\nu\Big]
\nonumber\\
&\hspace{1cm}+\delta_{\mu\nu}\sum\limits_{\substack{\alpha=1 \\ \alpha\neq\mu}}^4
\Big[K^{(sc)}_{\mu\alpha}(p,q)\gamma_\mu+K^{(cs)}_{\mu\alpha}(p,q)\Big]
\bigg)
\nonumber\\
+\ri \rho_3&\bigg(
\frac{4}{3}(1-\delta_{\mu\nu})
\sum\limits_{\substack{\rho=1 \\ \neq(\rho;\mu,\nu)}}^4 
\Big[L^{(csc)}_{\mu\nu\rho}(p,q,k_2)\gamma_\mu+L^{(scc)}_{\mu\nu\rho}(p,q,k_2)\gamma_\nu-L^{(sss)}_{\mu\nu\rho}(p,q,k_2)\gamma_\rho
\Big]\nonumber\\
&\hspace{1cm}+\frac{1}{3} \delta_{\mu\nu}
\sum\limits_{\substack{\alpha,\rho=1 \\ \neq(\alpha,\rho;\mu)}}^4
\Big[K^{(scc)}_{\mu\alpha\rho}(p,q)\gamma_\mu+2K^{(ccs)}_{\mu\alpha\rho}(p,q)\gamma_\rho\Big]
\bigg)
\nonumber\\
+\ri \rho_4&\bigg(
\frac{1}{3} (1-\delta_{\mu\nu})
\sum\limits_{\substack{\rho,\sigma=1 \\ \neq(\rho,\sigma,\mu,\nu)}}^4 
\Big[L^{(cscc)}_{\mu\nu\rho\sigma}(p,q,k_2)\gamma_\mu+L^{(sccc)}_{\mu\nu\rho\sigma}(p,q,k_2)\gamma_\nu-2L^{(sssc)}_{\mu\nu\rho\sigma}(p,q,k_2)\gamma_\rho\Big]
\nonumber\\
&\hspace{1cm}+\frac{1}{9} \delta_{\mu\nu}
\sum\limits_{\substack{\alpha,\rho,\sigma=1 \\ \neq(\alpha,\rho,\sigma,\mu)}}^4
\Big[K^{(sccc)}_{\mu\alpha\rho\sigma}(p,q)\gamma_\mu+3K^{(ccsc)}_{\mu\alpha\rho\sigma}(p,q)\gamma_\rho
\Big]\bigg)\Bigg\}
\end{align}
and needs to be complemented with the piece linear in $\csw^{(0)}=r$ that comes from the clover term.

\subsubsection{The $\mathbf{q\bar{q}ggg}$-vertex}

The $q\bar{q}ggg$ vertex for the Brillouin action takes the form
\begin{align}\label{eq:V3}
V_{3\,\mu\nu\rho}^{abc}&(p,q,k_1,k_2,k_3)=a^2g_0^3 T^aT^bT^c
\Bigg\{
\frac{1}{6}\delta_{\mu\nu}\delta_{\mu\rho}\big( r\lambda_1s(p_\mu+q_\mu)+ 2 \ri \rho_1 c(p_\mu+q_\mu)\gamma_\mu\big)
\nonumber\\
+r\lambda_2&\bigg(
\frac{1}{2}\delta_{\mu\nu}(1-\delta_{\mu\rho})
L^{(cs)}_{\mu\rho}(p,q,k_3)
+\frac{1}{2}\delta_{\nu\rho}(1-\delta_{\mu\nu})
L^{(sc)}_{\mu\nu}(p,q,k_2+k_3)
\nonumber\\
&+\frac{1}{6}\delta_{\mu\nu}\delta_{\mu\rho}
\sum\limits_{\substack{\alpha=1 \\ \alpha\neq\mu}}^4
K^{(sc)}_{\mu\alpha}(p,q)
\bigg)
\nonumber\\
+r\lambda_3&\bigg(
-\frac{2}{3}(1-\delta_{\mu\nu})(1-\delta_{\mu\rho})(1-\delta_{\nu\rho})
M^{(sss)}_{\mu\nu\rho}(p,q,k_2,k_3)
\nonumber\\
&+\frac{1}{3}\delta_{\mu\nu}(1-\delta_{\mu\rho})
\sum\limits_{\substack{\alpha=1 \\ \neq(\alpha,\mu,\rho)}}^4
L^{(csc)}_{\mu\rho\alpha}(p,q,k_3)
+\frac{1}{3}\delta_{\nu\rho}(1-\delta_{\mu\nu})
\sum\limits_{\substack{\alpha=1 \\ \neq(\alpha,\mu,\nu)}}^4
L^{(scc)}_{\mu\nu\alpha}(p,q,k_2+k_3)
\nonumber\\
&+\frac{1}{18} \delta_{\mu\nu}\delta_{\mu\rho}
\sum\limits_{\substack{\alpha,\beta=1 \\ \neq(\alpha,\beta,\mu)}}^4K^{(scc)}_{\mu\alpha\beta}(p,q)
\bigg)
\nonumber\\
+r\lambda_4&\bigg(
-\frac{1}{3}(1-\delta_{\mu\nu})(1-\delta_{\mu\rho})(1-\delta_{\nu\rho})
\sum\limits_{\substack{\sigma=1 \\ \neq(\sigma,\mu,\nu,\rho)}}^4
M^{(sssc)}_{\mu\nu\rho\sigma}(p,q,k_2,k_3)
\nonumber\\
&+\frac{1}{12}\delta_{\mu\nu}(1-\delta_{\nu\rho})
\sum\limits_{\substack{\alpha,\sigma=1 \\ \neq(\alpha,\sigma;\mu,\rho)}}^4
L^{(cscc)}_{\mu\rho\alpha\sigma}(p,q,k_3)
+\frac{1}{12}\delta_{\nu\rho}(1-\delta_{\mu\nu})
\sum\limits_{\substack{\alpha,\sigma=1 \\ \neq(\alpha,\sigma;\mu,\nu)}}^4
L^{(sccc)}_{\mu\nu\alpha\sigma}(p,q,k_2+k_3)
\nonumber\\
&+\frac{1}{54} \delta_{\mu\nu}\delta_{\mu\rho}
\sum\limits_{\substack{\alpha,\beta,\sigma=1 \\ \neq(\alpha,\beta,\sigma,\mu)}}^4
K^{(sccc)}_{\mu\alpha\beta\sigma}(p,q)
\bigg)
\nonumber\\
+\ri \rho_2&\bigg(
\frac{1}{3}\delta_{\mu\nu}\delta_{\mu\rho}
\sum\limits_{\substack{\alpha=1 \\ \alpha\neq\mu}}^4
\Big[K^{(cc)}_{\mu\alpha}(p,q)\gamma_\mu-K^{(ss)}_{\mu\alpha}(p,q)\gamma_\alpha\Big]
\nonumber\\
&+\delta_{\mu\nu}(1-\delta_{\mu\rho})
\Big[L^{(cc)}_{\mu\rho}(p,q,k_3)\gamma_\rho-L^{(ss)}_{\mu\rho}(p,q,k_3)\gamma_\mu\Big]
\nonumber\\
&+\delta_{\nu\rho}(1-\delta_{\mu\nu})
\Big[L^{(cc)}_{\mu\rho}(p,q,k_2+k_3)\gamma_\mu-L^{(ss)}_{\mu\rho}(p,q,k_2+k_3)\gamma_\nu\Big]
\bigg)
\nonumber\\
+\ri \rho_3&\bigg(
-\frac{4}{3}(1-\delta_{\mu\nu})(1-\delta_{\mu\rho})(1-\delta_{\nu\rho})
\nonumber\\&\times
\Big[M^{(css)}_{\mu\nu\rho}(p,q,k_2,k_3)\gamma_\mu+
M^{(scs)}_{\mu\nu\rho}(p,q,k_2,k_3)\gamma_\nu+
M^{(ssc)}_{\mu\nu\rho}(p,q,k_2,k_3)\gamma_\rho
\Big]
\nonumber\\
&+\frac{1}{9}\delta_{\mu\nu}\delta_{\mu\rho}
\sum\limits_{\substack{\alpha,\beta=1 \\ \neq(\alpha,\beta,\mu)}}^4 
\Big[K^{(ccc)}_{\mu\alpha\beta}(p,q)\gamma_\mu 
-K^{(ssc)}_{\mu\alpha\beta}(p,q)\gamma_\alpha
-K^{(scs)}_{\mu\alpha\beta}(p,q)\gamma_\beta\Big]
\nonumber\\
&-\frac{2}{3}\delta_{\mu\nu}(1-\delta_{\mu\rho})
\sum\limits_{\substack{\alpha=1 \\ \neq(\alpha,\mu,\rho)}}^4 
\Big[L^{(ssc)}_{\mu\rho\alpha}(p,q,k_3)\gamma_\mu
-L^{(ccc)}_{\mu\rho\alpha}(p,q,k_3)\gamma_\rho
+L^{(css)}_{\mu\rho\alpha}(p,q,k_3)\gamma_\alpha\Big]
\nonumber\\
&-\frac{2}{3}\delta_{\nu\rho}(1-\delta_{\mu\nu})
\sum\limits_{\substack{\alpha=1 \\ \neq(\alpha,\mu,\nu)}}^4 
\Big[-L^{(ccc)}_{\mu\nu\alpha}(p,q,k_2+k_3)\gamma_\mu
+L^{(ssc)}_{\mu\nu\alpha}(p,q,k_2+k_3)\gamma_\nu
\nonumber\\
&\hspace{4cm}
+L^{(scs)}_{\mu\nu\alpha}(p,q,k_2+k_3)\gamma_\alpha\Big]
\bigg)
\nonumber\\
+\ri \rho_4&\bigg(
-\frac{2}{3}(1-\delta_{\mu\nu})(1-\delta_{\mu\rho})(1-\delta_{\nu\rho})
\sum\limits_{\substack{\sigma=1 \\ \neq(\sigma,\mu,\nu,\rho)}}^4 
\Big[M^{(cssc)}_{\mu\nu\rho\sigma}(p,q,k_2,k_3)\gamma_\mu
\nonumber\\
&
+M^{(scsc)}_{\mu\nu\rho\sigma}(p,q,k_2,k_3)\gamma_\nu
+M^{(sscc)}_{\mu\nu\rho\sigma}(p,q,k_2,k_3)\gamma_\rho
-M^{(ssss)}_{\mu\nu\rho\sigma}(p,q,k_2,k_3)\gamma_\sigma
\Big]
\nonumber\\
&+\frac{1}{27} \delta_{\mu\nu}\delta_{\mu\rho}
\sum\limits_{\substack{\alpha,\beta,\sigma=1 \\ \neq(\alpha,\beta,\sigma,\mu)}}^4 
\Big[K^{(cccc)}_{\mu\alpha\beta\sigma}(p,q)\gamma_\mu 
-K^{(sscc)}_{\mu\alpha\beta\sigma}(p,q)\gamma_\alpha
\nonumber\\
&\hspace{3.3cm}
-K^{(scsc)}_{\mu\alpha\beta\sigma}(p,q)\gamma_\beta
-K^{(sccs)}_{\mu\alpha\beta\sigma}(p,q)\gamma_\sigma\Big]
\nonumber\\
&-\frac{1}{6}\delta_{\mu\nu}(1-\delta_{\mu\rho})
\sum\limits_{\substack{\alpha,\sigma=1 \\ \neq(\alpha,\sigma,\mu,\rho)}}^4 
\Big[L^{(sscc)}_{\mu\rho\alpha\sigma}(p,q,k_3)\gamma_\mu
-L^{(cccc)}_{\mu\rho\alpha\sigma}(p,q,k_3)\gamma_\rho
\nonumber\\
&\hspace{4.5cm}
+L^{(cssc)}_{\mu\rho\alpha\sigma}(p,q,k_3)\gamma_\alpha
+L^{(cscs)}_{\mu\rho\alpha\sigma}(p,q,k_3)\gamma_\sigma\Big]
\nonumber\\
&+\frac{1}{6}\delta_{\nu\rho}(1-\delta_{\mu\nu})
\sum\limits_{\substack{\alpha,\sigma=1 \\ \neq(\alpha,\sigma,\mu,\nu)}}^4 
\Big[L^{(cccc)}_{\mu\nu\alpha\sigma}(p,q,k_2+k_3)\gamma_\mu
-L^{(sscc)}_{\mu\nu\alpha\sigma}(p,q,k_2+k_3)\gamma_\nu
\nonumber\\
&\hspace{4.5cm}
+L^{(scsc)}_{\mu\nu\alpha\sigma}(p,q,k_2+k_3)\gamma_\alpha
+L^{(sccs)}_{\mu\nu\alpha\sigma}(p,q,k_2+k_3)\gamma_\sigma\Big]
\bigg)\Bigg\}
\end{align}
and needs to be complemented with the piece linear in $\csw^{(0)}=r$ that comes from the clover term.

\subsection{Feynman Rules of the Clover Term\label{app:clover_rules}}
Here we give the contributions to the $\bar{q}qgg$- and $\bar{q}qggg$-vertices coming from the clover term. Similar expressions can be found in Refs.~\cite{Aoki:2003sj,Horsley:2008ap}.
\begin{align}
V_{2c\,\mu\nu}^{ab}(p,q,k_1,k_2)&=
\ri\, a\,\csw^{(0)}g_0^2 T^aT^b \bigg(
\frac {1}{4}\delta_{\mu\nu}\sum\limits_\rho\sigma_{\mu\rho}s(q_\mu-p_\mu)[\bar{s}(k_{2\rho})-\bar{s}(k_{1\rho})]\nonumber\\
&+\sigma_{\mu\nu}\bigg[c(k_{1\nu})c(k_{2\mu})c(q_\mu-p_\mu)c(q_\nu-p_\nu)
-\frac{1}{2}c(k_{1\mu})c(k_{2\nu})\bigg]\bigg)
\label{clover_vertex_2}
\end{align}
\begin{align}
V_{3c\,\mu\nu\rho}^{abc}&(p,q,k_1,k_2,k_3)=
\ri\,a^2\,\csw^{(0)}\,g_0^3 T^aT^bT^c\Bigg(
 \delta_{\mu\nu}\delta_{\mu\rho}\sum\limits_\alpha\sigma_{\mu\alpha}\bigg[-\frac{1}{12}c(q_\mu-p_\mu)\bar{s}(q_\alpha-p_\alpha)\nonumber\\
&-\frac{1}{2}c(q_\mu-p_\mu)c(q_\alpha-p_\alpha)c(k_{3\alpha}-k_{1\alpha})s(k_{2\alpha})\bigg]\nonumber\\
&+\delta_{\mu\nu}\sigma_{\mu\rho}\bigg[-\frac{1}{2}c(q_\mu-p_\mu)c(q_\rho-p_\rho)c(k_{1\rho}+k_{2\rho})s(k_{3\mu})
+\frac{1}{4}s(k_{1\mu}+k_{2\mu})c(2k_{2\rho}+k_{3\rho})\bigg]\nonumber\\
&+\delta_{\nu\rho}\sigma_{\mu\nu}\bigg[-\frac{1}{2}c(q_\mu-p_\mu)c(q_\nu-p_\nu)c(k_{2\mu}+k_{3\mu})s(k_{1\nu})
+\frac{1}{4}s(k_{2\nu}+k_{3\nu})c(k_{1\mu}+2k_{2\mu})\bigg]\nonumber\\
&+\delta_{\mu\rho}\sigma_{\mu\nu}\bigg[\frac{1}{2}c(q_\nu-p_\nu)c(k_{3\nu}-k_{1\nu})s(k_{1\mu}+2k_{2\mu}+k_{3\mu})\bigg]\Bigg)
\label{clover_vertex_3}
\end{align}
%
\subsection{Feynman Rules of the Gauge Action\label{app:gauge_rules}}
The following gluon propagator and three-gluon vertex have been derived in Refs.~\cite{Weisz:1982zw,Weisz:1983bn}.
The gluon propagator in a general covariant gauge is
\begin{align}
G_{\mu\nu}(k)&=\frac{1}{4(s^2(k))^2}\bigg(\alpha s(k_\mu)s(k_\nu)+\sum\limits_\rho\big(\delta_{\mu\nu}s(k_\rho)-\delta_{\mu\rho}s(k_\rho)\big)s(k_\rho)A_{\rho\nu}(k)\bigg)
\end{align}
and we shall use the Feynman gauge with $\alpha=1$.
The symmetric matrix $A$ is given by
\begin{align}
A_{\mu\nu}(k)&=\frac{(1-\delta_{\mu\nu})}{\Delta(k)}\Bigg[(s^2(k))^2-4c_1s^2(k)\Big(2\sum\limits_\rho s(k_\rho)^4 +s^2(k)\sum\limits_{\rho\neq\mu,\nu}s(k_\rho)^2 \Big)\nonumber\\
&+16c_1^2\bigg(\Big(\sum\limits_\rho s(k_\rho)^4\Big)^2 
+s^2(k)\sum\limits_\rho s(k_\rho)^4\sum\limits_{\tau\neq\mu\nu}s(k_\tau)^2
+(s^2(k))^2\prod\limits_{\rho\neq\mu,\nu}s(k_\rho)^2\bigg)\Bigg]
\end{align}
with the denominator taking the form
\begin{align}
\Delta(k)&=\bigg(s^2(k)-4c_1\sum\limits_\rho s(k_\rho)^4\bigg)\Bigg[s^2(k)
-4c_1\Big((s^2(k))^2+\sum\limits_\tau s(k_\tau)^4\Big)\nonumber\\
&+8c_1^2\bigg((s^2(k))^3+2\sum\limits_\tau s(k_\tau)^6-s^2(k)\sum\limits_\tau 
s(k_\tau)^4\bigg)\Bigg]\nonumber\\
&-16c_1^3\sum\limits_\rho s(k_\rho)^4\prod\limits_{\tau\neq\rho}s(k_\tau)^2
\;.
\end{align}
Similarly, the three-gluon vertex depicted in Figure~\ref{fig:Vg3} is found to be
\begin{align}
V_{g3\,\mu\nu\rho}^{abc}(k_1,k_2,k_3)=-\frac{\ri g_0}{6}f^{abc}\Big(c_0V_{g3\,\mu\nu\rho}^{(0)}(k_1,k_2,k_3)+c_1V_{g3\,\mu\nu\rho}^{(1)}(k_1,k_2,k_3)\Big)
\end{align}
with
\begin{align}
V_{g3\,\mu\nu\rho}^{(0)}(k_1,k_2,k_3)&=2\Big[
\delta_{\mu\nu}s(k_{1\rho}-k_{2\rho})c(k_{3\mu})
+2\mathrm{\,cyclic\,perms\,}\Big]\\
V_{g3\,\mu\nu\rho}^{(1)}(k_1,k_2,k_3)&=
8 V_{g3\,\mu\nu\rho}^{(0)}(k_1,k_2,k_3)\nonumber\\
&+8\bigg[\delta_{\mu\nu}\Big(c(k_{3\mu})\Big[s(k_{1\mu}-k_{2\mu})(\delta_{\mu\rho}s^2(k_{3})-s(k_{3\mu})s(k_{3\rho}))\nonumber\\
&-s(k_{1\rho}-k_{2\rho})\big(s(k_{1\rho})^2+s(k_{2\rho}^2)\big)\Big]
\nonumber\\
&+s(k_{1\rho}-k_{2\rho})\big(s(k_{1\mu})s(k_{2\mu})-c(k_{1\mu})c(k_{2\mu})s(k_{3\mu})^2\big)\Big)\nonumber\\
&+2\mathrm{\,cyclic\,perms\,}\bigg]
\;.
\end{align}
\section{Example calculation of diagram (c)\label{app:dia_c}}

The integral corresponding to diagram (c) is
\begin{align}
\Lambda_\mu^{a(1)(c)}=
6\intd{k}  \sum\limits_{\nu,\rho,\sigma,\tau}\sum\limits_{b,c}
 & V^{b,c}_{2\,\nu\rho}(p,q,k,q-p-k)
 G_{\nu\sigma}(k)
 G_{\rho\tau}(p-q+k)\nonumber\\
\times& V^{abc}_{3g\, \mu\sigma\tau}(q-p,-k,p-q+k)
\;.
\end{align}
The pre-factor $6=2\cdot 3 $ is due to the possible permutations of the internal gluon lines.
The simplest form of this integral is in the case of the Wilson fermion action and the plaquette gluon action.
In order to compute $G_1^{(c)}$ by Eq.~(\ref{eq:G1}), we need the derivatives
\begin{align}
\frac{\partial}{\partial p_\nu}\Lambda_\mu^{a(1)(c)}\bigg|_{p,q=0}^{\mu\neq\nu}
=&g_0^3 N_cT^a\intdk \Bigg\{\frac{-3\bar{s}(k_\mu)\bar{s}(k_\nu)+\ri  s^2(k)\bar{s}(k_\mu)\gamma_\nu}{64(s^2(k))^3}
\nonumber \\
&-\ri\csw^{(0)}\cdot\frac{3c(k_\mu)^2\bar{c}(k_\nu)\sigma_{\mu\nu}+\bar{s}(k_\mu)\sum_\rho\sigma_{\nu\rho}\bar{s}(k_\rho)}{64(s^2(k))^2}\Bigg\}
\\
\frac{\partial}{\partial q_\nu}\Lambda_\mu^{a(1)(c)}\bigg|_{p,q=0}^{\mu\neq\nu}
=&g_0^3 N_cT^a\intdk \Bigg\{\frac{3\bar{s}(k_\mu)\bar{s}(k_\nu)+\ri  s^2(k)\bar{s}(k_\mu)\gamma_\nu}{64(s^2)^3}
\nonumber \\
&+\ri\csw^{(0)}\cdot\frac{3c(k_\mu)^2\bar{c}(k_\nu)\sigma_{\mu\nu}+\bar{s}(k_\mu)\sum_\rho\sigma_{\nu\rho}\bar{s}(k_\rho)}{64(s^2(k))^2}\Bigg\}
\\
\frac{\partial}{\partial p_\mu}\Lambda_\mu^{a(1)(c)}\bigg|_{p,q=0}
=&g_0^3 N_cT^a\intdk \Bigg\{\frac{3\bar{s}(k_\mu)^2-3\bar{c}(k_\mu)s^2}{64(s^2(k))^3}
-\ri \csw^{(0)}\frac{\bar{s}(k_\mu)\sum_\rho\sigma_{\mu\rho}\bar{s}(k_\rho)}{64(s^2(k))^2}\Bigg\}
\\
\frac{\partial}{\partial q_\mu}\Lambda_\mu^{a(1)(c)}\bigg|_{p,q=0}
=&-\frac{\partial}{\partial p_\mu}\Lambda_\mu^{a(1)(c)}\bigg|_{p,q=0}\;.
\end{align}
Here we may use that under the integral only even functions will survive, therefore terms of the form $s(k_\mu)s(k_\nu)$ and $\bar{s}(k_\mu)\bar{s}(k_\nu)$ with $\mu\neq\nu$ do not contribute.
This yields
\begin{align}
&g_0^3T^aG_1^{(c)}=-\frac{1}{8} \text{Tr}\bigg[- \bigg(\frac{\partial}{\partial p_\nu}-\frac{\partial}{\partial q_\nu}\bigg)\Lambda^{a(1)}_\mu\gamma_\nu\gamma_\mu  \bigg]^{\mu\neq\nu}_{p,q\rightarrow 0}\\
&=-\frac{1}{8} g_0^3 N_cT^a\intdk\text{Tr}\bigg[  \bigg(\frac{3\bar{s}(k_\mu)\bar{s}(k_\nu)}{32(s^2(k))^3} 
+\ri \csw^{(0)}\frac{(3c(k_\mu)^2\bar{c}(k_\nu)-\bar{s}(k_\mu)^2)\sigma_{\mu\nu}}{32(s^2(k))^2} \bigg)\gamma_\nu\gamma_\mu\bigg]\\
&=g_0^3 N_c \csw^{(0)} T^a\intdk \frac{3c(k_\mu)^2\bar{c}(k_\nu)-\bar{s}(k_\mu)^2}{64(s^2(k))^2}
\end{align}
and this integral is divergent and cannot be directly calculated numerically.
In order to regularize it we subtract the simple lattice integral $\mathcal{B}_2$
\begin{align}
\intdk \bigg(\frac{3c(k_\mu)^2\bar{c}(k_\nu)-\bar{s}(k_\mu)^2}{64(s^2(k))^2} - c\cdot \frac{1}{(4s^2(k))^2}\bigg)=\mathrm{finite}
\end{align}
where all we need to know is the appropriate coefficient $c$.
We can determine it easily by restoring factors of the lattice spacing $k\to ak$, expanding the first term to lowest order in $a$
\begin{align}
\intdk \frac{3}{4}\frac{1}{(k_1^2+k_2^2+k_3^2+k_4^2)^2}+\mathcal{O}(a^2)
\end{align}
and reading off $c=3/4$.
Thus the contribution to $\csw^{(1)}$ coming from diagram (c) is
\begin{align}
\csw^{(1)(c)}&=2G_1^{(c)}=\csw^{(0)}N_c\bigg(\frac{3}{2}\mathcal{B}_2(\mu)-0.041375026(1)\bigg)\\
&=\frac{9}{2}\mathcal{B}_2(\mu)-0.124125079(1)
\end{align}
where we have set $\csw^{(0)}=r=1$ and $N_c=3$ in the last step.
\section{Self-Energy as a function of $r$ \label{app:self_energy}}
%
To complement the results presented in Section~\ref{sec:self_energy}, Table~\ref{tab:sigma0_vs_r} shows our results
for the self-energy $\Sigma_0/a$ for eleven values of the Wilson parameter $r$ between $0.5$ and $1.5$.
The four options of the overall action (Wilson/Brillouin fermions on plaquette/L\"uscher-Weisz glue)
are given with a split-up between the tadpole and sunset contributions and the combination of the two diagrams.
\begin{table}[h]
\centering
\begin{tabular}{|c|c|c|c|c|c|c|}
\hline
\multicolumn{1}{|c|}{}&\multicolumn{2}{|c|}{$\Sigma_0^{\mathrm{(tadpole)}}$} &\multicolumn{2}{|c|}{$\Sigma_0^{\mathrm{(sunset)}}$}&\multicolumn{2}{|c|}{$\Sigma_0$}\\
\hline
$r$ & Wil./Plaq. & Bri./Plaq. & Wil./Plaq. & Bri./Plaq.& Wil./Plaq. & Bri./Plaq. \\
\hline
$0.5$ & $-24.4661(1)$ & $-24.4661(1)$ & $ 3.7505(1)$ & $ 4.1566(1)$ & $-20.7156(1)$ & $-20.3094(1)$ \\
$0.6$ & $-29.3593(1)$ & $-29.3593(1)$ & $ 6.3797(1)$ & $ 7.0854(1)$ & $-22.9797(1)$ & $-22.2739(1)$ \\
$0.7$ & $-34.2525(1)$ & $-34.2525(1)$ & $ 9.0474(1)$ & $ 9.8887(1)$ & $-25.2052(1)$ & $-24.3638(1)$ \\
$0.8$ & $-39.1458(1)$ & $-39.1458(1)$ & $11.7083(1)$ & $12.5913(1)$ & $-27.4375(1)$ & $-26.5545(1)$ \\
$0.9$ & $-44.0390(1)$ & $-44.0390(1)$ & $14.3433(1)$ & $15.2139(1)$ & $-29.6956(1)$ & $-28.8251(1)$ \\
$1.0$ & $-48.9322(1)$ & $-48.9322(1)$ & $16.9458(1)$ & $17.7727(1)$ & $-31.9864(1)$ & $-31.1595(1)$ \\
$1.1$ & $-53.8254(1)$ & $-53.8254(1)$ & $19.5147(1)$ & $20.2804(1)$ & $-34.3108(1)$ & $-33.5450(1)$ \\
$1.2$ & $-58.7186(1)$ & $-58.7186(1)$ & $22.0517(1)$ & $22.7465(1)$ & $-36.6669(1)$ & $-35.9721(1)$ \\
$1.3$ & $-63.6119(1)$ & $-63.6119(1)$ & $24.5597(1)$ & $25.1785(1)$ & $-39.0522(1)$ & $-38.4333(1)$ \\
$1.4$ & $-68.5051(1)$ & $-68.5051(1)$ & $27.0415(1)$ & $27.5823(1)$ & $-41.4636(1)$ & $-40.9227(1)$ \\
$1.5$ & $-73.3983(1)$ & $-73.3983(1)$ & $29.5000(1)$ & $29.9625(1)$ & $-43.8983(1)$ & $-43.4358(1)$ \\
\hline
\hline
$r$ & Wil./Sym. & Bri./Sym. & Wil./Sym. & Bri./Sym.& Wil./Sym. & Bri./Sym. \\
\hline
$0.5$ & $-20.2588(1)$ & $-19.5499(1)$ & $4.7572(1)$  & $ 4.6636(1)$ & $-15.5016(1)$ & $-14.8863(1)$ \\
$0.6$ & $-24.3106(1)$ & $-23.4599(1)$ & $7.1298(1)$  & $ 7.1531(1)$ & $-17.1808(1)$ & $-16.3068(1)$ \\
$0.7$ & $-28.3624(1)$ & $-27.3699(1)$ & $9.5352(1)$  & $ 9.5532(1)$ & $-18.8272(1)$ & $-17.817(1)$  \\
$0.8$ & $-32.4142(1)$ & $-31.2799(1)$ & $11.9382(1)$ & $11.8810(1)$ & $-20.4760(1)$ & $-19.399(1)$  \\
$0.9$ & $-36.4659(1)$ & $-35.1898(1)$ & $14.3235(1)$ & $14.151(1)$  & $-22.1425(1)$ & $-21.04(1)$   \\
$1.0$ & $-40.5178(1)$ & $-39.0998(1)$ & $16.6854(1)$ & $16.374(1)$  & $-23.8323(1)$ & $-22.73(1)$   \\
$1.1$ & $-44.5695(1)$ & $-43.0098(1)$ & $19.0229(1)$ & $18.560(1)$  & $-25.5467(1)$ & $-24.45(1)$   \\
$1.2$ & $-48.6213(1)$ & $-46.9198(1)$ & $21.3369(1)$ & $20.716(1)$  & $-27.2844(1)$ & $-26.20(1)$   \\
$1.3$ & $-52.6731(1)$ & $-50.8298(1)$ & $23.6293(1)$ & $22.846(1)$  & $-29.0438(1)$ & $-27.98(1)$   \\
$1.4$ & $-56.7248(1)$ & $-54.7397(1)$ & $25.9021(1)$ & $24.956(1)$  & $-30.8228(1)$ & $-29.78(1)$   \\
$1.5$ & $-60.7766(1)$ & $-58.6497(1)$ & $28.1575(1)$ & $27.048(1)$  & $-32.6191(1)$ & $-31.60(1)$   \\
\hline
\end{tabular}
\caption{Self energy $\Sigma_0$ of Wilson and Brillouin fermions for $N_c=3$ with plaquette and tree-level Symanzik improved (``L\"uscher-Weisz'') gauge action as a function of $r$.}
\label{tab:sigma0_vs_r}
\end{table}
%
\section{Results for $\csw^{(1)}$ as a function of $r$ \label{app:csw_r}}
%
To complement the results presented in Section~\ref{sec:csw1}, Table~\ref{tab:csw_final} shows our results
for the one-loop clover coefficient $\csw^{(1)}$ for eleven values of the Wilson parameter $r$ between $0.5$ and $1.5$.
The physical number of colours, $N_c=3$, has been plugged in.
The four columns encode for the overall action (Wilson/Brillouin fermions on plaquette/L\"uscher-Weisz glue), respectively.
\begin{table}[h]
\centering
\begin{tabular}{|c| l | l | l | l |}
\hline
\multicolumn{1}{|c|}{}& \multicolumn{4}{|c|}{$\csw^{(1)}$ ($N_c=3$)}\\
\hline
$r$ & \multicolumn{1}{|c|}{Wilson/Plaq.} & \multicolumn{1}{|c|}{Brillouin/Plaq.} & \multicolumn{1}{|c|}{Wilson/Sym.} & \multicolumn{1}{|c|}{Brillouin/Sym.} \\
\hline
$0.5$ &  $0.15603501(1)$ & $0.064568(1)$   & $0.1121103(1)$  & $0.045279(1)$ \\
$0.6$ &  $0.17891256(1)$ & $0.076957(1)$   & $0.1292256(1)$  & $0.054428(1)$ \\
$0.7$ &  $0.20141780(1)$ & $0.089006(1)$   & $0.14607072(1)$ & $0.063298(1)$ \\
$0.8$ &  $0.2237942(1) $ & $0.100777(1)$   & $0.1628066(1)$  & $0.071932(1)$ \\
$0.9$ &  $0.2461643(1) $ & $0.123626(1)$   & $0.1795168(1)$  & $0.080360(1)$ \\
$1.0$ &  $0.26858825(1)$ & $0.12362580(1)$ & $0.1962445(1)$  & $0.088601(1)$ \\
$1.1$ &  $0.29109341(1)$ & $0.13474599(1)$ & $0.21301078(1)$ & $0.096670(1)$ \\
$1.2$ &  $0.31369009(1)$ & $0.14568094(1)$ & $0.22982530(1)$ & $0.104577(1)$ \\
$1.3$ &  $0.33637959(1)$ & $0.15643861(1)$ & $0.2466912(1)$  & $0.112330(1)$ \\
$1.4$ &  $0.35915870(1)$ & $0.16702454(1)$ & $0.2636083(1)$  & $0.119934(1)$ \\
$1.5$ &  $0.38202196(1)$ & $0.17744258(1)$ & $0.2805746(1)$  & $0.127392(1)$ \\
\hline
\end{tabular}
\caption{Coefficient $\csw^{(1)}$ for Wilson and Brillouin fermions for $N_c=3$ as a function of $r$.}
\label{tab:csw_final}
\end{table}

To enable the reader to adjust the result to arbitrary $N_c$, Table~\ref{tab:csw_ncsplitup} gives the parts $\propto N_c$ and $\propto 1/N_c$ separately.
Plugging $N_c=3$ reproduces the results listed in Table~\ref{tab:csw_final}.
For future reference the splitup into contributions per diagram (for $N_c=3$) is given in Table~\ref{tab:csw_perdiagram_wilson} for Wilson fermions
and in Table~\ref{tab:csw_perdiagram_brillouin} for Brillouin fermions, respectively.
\newpage
\begin{table}[h]
\centering
\setlength\tabcolsep{4pt}
\begin{tabular}{|c| l | l | l | l |}
\hline
\multicolumn{1}{|c|}{} & \multicolumn{2}{|c|}{$N_c$} & \multicolumn{2}{|c|}{$1/N_c$} \\
\hline
$r$ & \multicolumn{1}{|c|}{Wilson/Plaq.} & \multicolumn{1}{|c|}{Brillouin/Plaq.} &\multicolumn{1}{|c|}{Wilson/Plaq.} & \multicolumn{1}{|c|}{Brillouin/Plaq.}\\
\hline
$0.5$ & $ 0.05715043(1)$ & $0.023738(1)$   & $-0.046248854(1)$  & $-0.0199408(1)$  \\
$0.6$ & $ 0.0656271(1)$  & $0.028384(1)$   & $-0.0539062426(2)$ & $-0.0245934(1)$  \\
$0.7$ & $ 0.07396653(1)$ & $0.032891(1)$   & $-0.06144543(1)$   & $-0.02900721(1)$ \\
$0.8$ & $ 0.08225622(1)$ & $0.037284(1)$   & $-0.0689235(1)$    & $-0.0332272(1)$  \\
$0.9$ & $ 0.09054080(1)$ & $0.041578(1)$   & $-0.07637429(1)$   & $-0.037282(1)$   \\
$1.0$ & $ 0.09884247(1)$ & $0.04578551(1)$ & $-0.08381749(1)$   & $-0.04119225(1)$ \\
$1.1$ & $ 0.10717161(1)$ & $0.049912(1)$   & $-0.09126427(1)$   & $-0.0449695(1)$  \\
$1.2$ & $ 0.1155323(1)$  & $0.053963(1)$   & $-0.0987207(1)$    & $-0.04862324(1)$ \\
$1.3$ & $ 0.1239253(1)$  & $0.057942(1)$   & $-0.1061899(1)$    & $-0.05215987(1)$ \\
$1.4$ & $ 0.1323498(1)$  & $0.061850(1)$   & $-0.1136732(1)$    & $-0.05558402(1)$ \\
$1.5$ & $ 0.1408040(1)$  & $0.065691(1)$   & $-0.1211708(2)$    & $-0.0588991(1)$ \\
\hline
\hline
$r$ & \multicolumn{1}{|c|}{Wilson/Sym.} & \multicolumn{1}{|c|}{Brillouin/Sym.} & \multicolumn{1}{|c|}{Wilson/Sym.} & \multicolumn{1}{|c|}{Brillouin/Sym.}\\
\hline
$0.5$ & $0.04087367(1)$ & $0.016533(1)$ & $-0.0315321(1)$   & $-0.0129631(1)$  \\
$0.6$ & $0.04718026(1)$ & $0.019944(1)$ & $-0.0369456(1)$   & $-0.0162088(1)$  \\
$0.7$ & $0.05338807(1)$ & $0.023240(1)$ & $-0.04228046(1)$  & $-0.01926933(1)$ \\
$0.8$ & $0.0595545(1)$  & $0.026441(1)$ & $-0.04757044(1)$  & $-0.02217492(1)$ \\
$0.9$ & $0.0657097(1)$  & $0.029558(1)$ & $-0.05283676(1)$  & $-0.0249463(1)$  \\
$1.0$ & $0.0718695(1)$  & $0.032600(1)$ & $-0.05809245(1)$  & $-0.0275974(1)$  \\
$1.1$ & $0.07804196(1)$ & $0.035572(1)$ & $-0.063345328(1)$ & $-0.03013818(1)$ \\
$1.2$ & $0.08423066(1)$ & $0.038478(1)$ & $-0.06860003(1)$  & $-0.03257563(1)$ \\
$1.3$ & $0.09043698(1)$ & $0.041323(1)$ & $-0.0738592(1)$   & $-0.03491483(1)$ \\
$1.4$ & $0.09666103(1)$ & $0.044107(1)$ & $-0.0791243(1)$   & $-0.0371596(1)$  \\
$1.5$ & $0.1029022(1)$  & $0.046832(1)$ & $-0.08439603(1)$  & $-0.0393127(1)$  \\
\hline
\end{tabular}
\caption{Parts $\propto N_c$ and $\propto 1/N_c$ of $\csw^{(1)}$ for Wilson and Brillouin fermions as a function of $r$.}
\label{tab:csw_ncsplitup}
\end{table}
\newpage
\pagestyle{empty}
\begin{table}[!p]
\centering
\setlength\tabcolsep{3pt}
\begin{tabular}{|c|r l |r l |r l | r l |}
\hline
\multicolumn{7}{|c|}{Wilson/Plaq.} \\
\hline
$r$ & \multicolumn{2}{|c|}{(a)}& \multicolumn{2}{|c|}{(b)}& \multicolumn{2}{|c|}{(c)} \\
\hline
$0.5$ & $-1/6  \mathcal{B}_2$ & $+0.0048466288(1)$ & $-9/4   \mathcal{B}_2$ & $+0.066223184(1)$  & $9/4   \mathcal{B}_2$ & $-0.062062540(1)$ \\
$0.6$ & $-1/5  \mathcal{B}_2$ & $+0.005811036(1)$  & $-27/10 \mathcal{B}_2$ & $+0.078124152(1)$  & $27/10 \mathcal{B}_2$ & $-0.074475047(1)$ \\
$0.7$ & $-7/30 \mathcal{B}_2$ & $+0.0068019364(1)$ & $-63/20 \mathcal{B}_2$ & $+0.090021110(1)$  & $63/20 \mathcal{B}_2$ & $-0.086887555(1)$ \\
$0.8$ & $-4/15 \mathcal{B}_2$ & $+0.007810019(2)$  & $-18/5  \mathcal{B}_2$ & $+0.1019435380(1)$ & $18/5  \mathcal{B}_2$ & $-0.099300063(1)$ \\
$0.9$ & $-3/10 \mathcal{B}_2$ & $+0.00882832(2)$   & $-81/20 \mathcal{B}_2$ & $+0.113901379(1)$  & $81/20 \mathcal{B}_2$ & $-0.11171257(1) $ \\
$1.0$ & $-1/3  \mathcal{B}_2$ & $+0.009852153(1)$  & $-9/2   \mathcal{B}_2$ & $+0.125895883(1)$  & $9/2   \mathcal{B}_2$ & $-0.124125079(1)$ \\
$1.1$ & $-11/30\mathcal{B}_2$ & $+0.010878501(1)$  & $-99/20 \mathcal{B}_2$ & $+0.13792469(1)$   & $99/20 \mathcal{B}_2$ & $-0.136537587(1)$ \\
$1.2$ & $-2/5  \mathcal{B}_2$ & $+0.011905489(1)$  & $-27/5  \mathcal{B}_2$ & $+0.149984155(1)$  & $27/5  \mathcal{B}_2$ & $-0.148950095(1)$ \\
$1.3$ & $-13/30\mathcal{B}_2$ & $+0.012931987(1)$  & $-117/20\mathcal{B}_2$ & $+0.16207045(1)$   & $117/20\mathcal{B}_2$ & $-0.161362603(1)$ \\
$1.4$ & $-7/15 \mathcal{B}_2$ & $+0.013957336(1)$  & $-63/10 \mathcal{B}_2$ & $+0.174179964(1)$  & $63/10 \mathcal{B}_2$ & $-0.17377511(1) $ \\
$1.5$ & $-1/2  \mathcal{B}_2$ & $+0.014981183(1)$  & $-27/4  \mathcal{B}_2$ & $+0.186309485(1)$  & $27/4  \mathcal{B}_2$ & $-0.18618762(1) $ \\
\hline
$r$ & \multicolumn{2}{|c|}{(d)}& \multicolumn{2}{|c|}{(e)}& \multicolumn{2}{|c|}{(f)} \\
\hline
$0.5$ & \multicolumn{2}{|c|}{$0.148697267(1)$} & $1/12 \mathcal{B}_2$ & $-0.000834765(1)$ & $1/12 \mathcal{B}_2$ & $-0.000834765(1)$ \\
$0.6$ & \multicolumn{2}{|c|}{$0.178436720(1)$} & $1/10 \mathcal{B}_2$ & $-0.00449215(1)$  & $1/10 \mathcal{B}_2$ & $-0.00449215(1)$  \\
$0.7$ & \multicolumn{2}{|c|}{$0.208176174(1)$} & $7/60 \mathcal{B}_2$ & $-0.00834693(1)$  & $7/60 \mathcal{B}_2$ & $-0.00834693(1)$  \\
$0.8$ & \multicolumn{2}{|c|}{$0.237915627(1)$} & $2/15 \mathcal{B}_2$ & $-0.01228747(1)$  & $2/15 \mathcal{B}_2$ & $-0.01228747(1)$  \\
$0.9$ & \multicolumn{2}{|c|}{$0.267655081(1)$} & $3/20 \mathcal{B}_2$ & $-0.01625395(1)$  & $3/20 \mathcal{B}_2$ & $-0.01625395(1)$  \\
$1.0$ & \multicolumn{2}{|c|}{$0.297394534(1)$} & $1/6  \mathcal{B}_2$ & $-0.020214623(1)$ & $1/6  \mathcal{B}_2$ & $-0.020214623(1)$ \\
$1.1$ & \multicolumn{2}{|c|}{$0.327133988(1)$} & $11/60\mathcal{B}_2$ & $-0.02415309(1)$  & $11/60\mathcal{B}_2$ & $-0.02415309(1)$  \\
$1.2$ & \multicolumn{2}{|c|}{$0.356873441(2)$} & $1/5  \mathcal{B}_2$ & $-0.02806145(1)$  & $1/5  \mathcal{B}_2$ & $-0.02806145(1)$  \\
$1.3$ & \multicolumn{2}{|c|}{$0.386612895(2)$} & $13/60\mathcal{B}_2$ & $-0.03193657(1)$  & $13/60\mathcal{B}_2$ & $-0.03193657(1)$  \\
$1.4$ & \multicolumn{2}{|c|}{$0.416352348(2)$} & $7/30 \mathcal{B}_2$ & $-0.03577792(1)$  & $7/30 \mathcal{B}_2$ & $-0.03577792(1)$  \\
$1.5$ & \multicolumn{2}{|c|}{$0.446091802(2)$} & $1/4  \mathcal{B}_2$ & $-0.03958640(1)$  & $1/4  \mathcal{B}_2$ & $-0.03958640(1)$  \\
\hline
%
\multicolumn{7}{|c|}{Wilson/Sym.} \\
\hline
$r$ & \multicolumn{2}{|c|}{(a)}& \multicolumn{2}{|c|}{(b)}& \multicolumn{2}{|c|}{(c)} \\
\hline
$0.5$ & $-1/6  \mathcal{B}_2$ & $+0.005143110(1)$  & $-9/4   \mathcal{B}_2$ & $+0.065608(1)$    & $9/4   \mathcal{B}_2$ & $-0.06688903(1)$ \\
$0.6$ & $-1/5  \mathcal{B}_2$ & $+0.00618956(1)$   & $-27/10 \mathcal{B}_2$ & $+0.0781430(1)$   & $27/10 \mathcal{B}_2$ & $-0.08026684(1)$ \\
$0.7$ & $-7/3  \mathcal{B}_2$ & $+0.00725233(1)$   & $-63/20 \mathcal{B}_2$ & $+0.09070162(1)$  & $63/20 \mathcal{B}_2$ & $-0.0936446(1)$  \\
$0.8$ & $-4/15 \mathcal{B}_2$ & $+0.00832513(1)$   & $-18/5  \mathcal{B}_2$ & $+0.103291488(1)$ & $18/5  \mathcal{B}_2$ & $-0.1070225(1)$  \\
$0.9$ & $-3/10 \mathcal{B}_2$ & $+0.00940336(1)$   & $-81/20 \mathcal{B}_2$ & $+0.11591188(1)$  & $81/20 \mathcal{B}_2$ & $-0.12040026(1)$ \\
$1.0$ & $-1/3  \mathcal{B}_2$ & $+0.01048401(1)$   & $-9/2   \mathcal{B}_2$ & $+0.1285594(1)$   & $9/2   \mathcal{B}_2$ & $-0.1337781(1) $ \\
$1.1$ & $-11/30\mathcal{B}_2$ & $+0.011565172(1)$  & $-99/20 \mathcal{B}_2$ & $+0.1412300(1)$   & $99/20 \mathcal{B}_2$ & $-0.14715587(1)$ \\
$1.2$ & $-2/5  \mathcal{B}_2$ & $+0.0126457266(1)$ & $-27/5  \mathcal{B}_2$ & $+0.15391985(1)$  & $27/5  \mathcal{B}_2$ & $-0.1605337(1) $ \\
$1.3$ & $-13/30\mathcal{B}_2$ & $+0.01372504(1)$   & $-117/20\mathcal{B}_2$ & $+0.16662571(1)$  & $117/20\mathcal{B}_2$ & $-0.1739115(1) $ \\
$1.4$ & $-7/15 \mathcal{B}_2$ & $+0.01480276(1)$   & $-63/10 \mathcal{B}_2$ & $+0.17934475(1)$  & $63/10 \mathcal{B}_2$ & $-0.1872893(1) $ \\
$1.5$ & $-1/2  \mathcal{B}_2$ & $+0.0158788(1)$    & $-27/4  \mathcal{B}_2$ & $+0.19207465(1)$  & $27/4  \mathcal{B}_2$ & $-0.2006671(1) $ \\
\hline
$r$ & \multicolumn{2}{|c|}{(d)}& \multicolumn{2}{|c|}{(e)}& \multicolumn{2}{|c|}{(f)} \\
\hline
$0.5$ & \multicolumn{2}{|c|}{$0.11771939(1)$} & $1/12 \mathcal{B}_2$ & $-0.004735883(1)$ & $1/12 \mathcal{B}_2$ & $-0.004735883(1)$ \\
$0.6$ & \multicolumn{2}{|c|}{$0.1412633(1)$}  & $1/10 \mathcal{B}_2$ & $-0.008051702(1)$ & $1/10 \mathcal{B}_2$ & $-0.008051702(1)$ \\
$0.7$ & \multicolumn{2}{|c|}{$0.1648071(1)$}  & $7/60 \mathcal{B}_2$ & $-0.011522869(1)$ & $7/60 \mathcal{B}_2$ & $-0.011522869(1)$ \\
$0.8$ & \multicolumn{2}{|c|}{$0.18835102(1)$} & $2/15 \mathcal{B}_2$ & $-0.01506930(1)$  & $2/15 \mathcal{B}_2$ & $-0.01506930(1)$  \\
$0.9$ & \multicolumn{2}{|c|}{$0.21189490(1)$} & $3/20 \mathcal{B}_2$ & $-0.01864652(1)$  & $3/20 \mathcal{B}_2$ & $-0.01864652(1)$  \\
$1.0$ & \multicolumn{2}{|c|}{$0.2354388(1)$}  & $1/6  \mathcal{B}_2$ & $-0.022229808(1)$ & $1/6  \mathcal{B}_2$ & $-0.022229808(1)$ \\
$1.1$ & \multicolumn{2}{|c|}{$0.25898265(1)$} & $11/60\mathcal{B}_2$ & $-0.0258056(1)$   & $11/60\mathcal{B}_2$ & $-0.0258056(1)$   \\
$1.2$ & \multicolumn{2}{|c|}{$0.2825265(1)$}  & $1/5  \mathcal{B}_2$ & $-0.0293666(1)$   & $1/5  \mathcal{B}_2$ & $-0.0293666(1)$   \\
$1.3$ & \multicolumn{2}{|c|}{$0.3060704(1)$}  & $13/60\mathcal{B}_2$ & $-0.032909247(1)$ & $13/60\mathcal{B}_2$ & $-0.032909247(1)$ \\
$1.4$ & \multicolumn{2}{|c|}{$0.3296143(1)$}  & $7/30 \mathcal{B}_2$ & $-0.0364321(1)$   & $7/30 \mathcal{B}_2$ & $-0.0364321(1)$   \\
$1.5$ & \multicolumn{2}{|c|}{$0.3531582(1)$}  & $1/4  \mathcal{B}_2$ & $-0.0399350(1)$   & $1/4  \mathcal{B}_2$ & $-0.0399350(1)$   \\
\hline
\end{tabular}
\caption{Contributions to ${\csw^{(1)}}_\mathrm{Wilson}$ for $N_c=3$ from each diagram as a function of $r$.}
\label{tab:csw_perdiagram_wilson}
\end{table}

\newpage

\begin{table}[!p]
\setlength\tabcolsep{3pt}
\begin{tabular}{|c|r l |r l |r l | r l |}
\hline
\multicolumn{7}{|c|}{Brillouin/Plaq.} \\
\hline
$r$ & \multicolumn{2}{|c|}{(a)}& \multicolumn{2}{|c|}{(b)}& \multicolumn{2}{|c|}{(c)} \\
\hline
$0.5$ & $-1/6  \mathcal{B}_2$ & $+0.00455393(1)$  & $-9/4   \mathcal{B}_2$ & $+0.052949245(1)$ & $9/4   \mathcal{B}_2$ & $-0.050279429(1)$  \\
$0.6$ & $-1/5  \mathcal{B}_2$ & $+0.005634969(1)$ & $-27/10 \mathcal{B}_2$ & $+0.061774874(1)$ & $27/10 \mathcal{B}_2$ & $-0.06033531(1)$   \\
$0.7$ & $-7/30 \mathcal{B}_2$ & $+0.00673256(1)$  & $-63/20 \mathcal{B}_2$ & $+0.07076411(1) $ & $63/20 \mathcal{B}_2$ & $-0.07039120(1)$   \\
$0.8$ & $-4/15 \mathcal{B}_2$ & $+0.00783569(1)$  & $-18/5  \mathcal{B}_2$ & $+0.07987858(1) $ & $18/5  \mathcal{B}_2$ & $-0.0804470863(1)$ \\
$0.9$ & $-3/10 \mathcal{B}_2$ & $+0.00893906(1)$  & $-81/20 \mathcal{B}_2$ & $+0.08908850(1) $ & $81/20 \mathcal{B}_2$ & $-0.0905029720(1)$ \\
$1.0$ & $-1/3  \mathcal{B}_2$ & $+0.010040221(1)$ & $-9/2   \mathcal{B}_2$ & $+0.098371668(1)$ & $9/2   \mathcal{B}_2$ & $-0.100558858(1)$  \\
$1.1$ & $-11/30\mathcal{B}_2$ & $+0.01113809(1)$  & $-99/20 \mathcal{B}_2$ & $+0.107711629(1)$ & $99/20 \mathcal{B}_2$ & $-0.1106147437(1)$ \\
$1.2$ & $-2/5  \mathcal{B}_2$ & $+0.012232304(1)$ & $-27/5  \mathcal{B}_2$ & $+0.117096140(1)$ & $27/5  \mathcal{B}_2$ & $-0.12352029(1)$   \\
$1.3$ & $-13/30\mathcal{B}_2$ & $+0.013322836(1)$ & $-117/20\mathcal{B}_2$ & $+0.126515999(1)$ & $117/20\mathcal{B}_2$ & $-0.13072652(1)$   \\
$1.4$ & $-7/15 \mathcal{B}_2$ & $+0.01440984(1)$  & $-63/10 \mathcal{B}_2$ & $+0.135964212(1)$ & $63/10 \mathcal{B}_2$ & $-0.14078240(1)$   \\
$1.5$ & $-1/2  \mathcal{B}_2$ & $+0.01549358(1)$  & $-27/4  \mathcal{B}_2$ & $+0.145435401(1)$ & $27/4  \mathcal{B}_2$ & $-0.150838287(1)$  \\
\hline
$r$ & \multicolumn{2}{|c|}{(d)}& \multicolumn{2}{|c|}{(e)}& \multicolumn{2}{|c|}{(f)} \\
\hline
$0.5$ & \multicolumn{2}{|c|}{$0.073247934(1)$} & $1/12 \mathcal{B}_2$ & $-0.007951680(1)$  & $1/12 \mathcal{B}_2$ & $-0.007951680(1)$  \\
$0.6$ & \multicolumn{2}{|c|}{$0.087413353(1)$} & $1/10 \mathcal{B}_2$ & $-0.00876542(1)$   & $1/10 \mathcal{B}_2$ & $-0.00876542(1)$   \\
$0.7$ & \multicolumn{2}{|c|}{$0.101417385(1)$} & $7/60 \mathcal{B}_2$ & $-0.0097580591(1)$ & $7/60 \mathcal{B}_2$ & $-0.0097580591(1)$ \\
$0.8$ & \multicolumn{2}{|c|}{$0.115260026(1)$} & $2/15 \mathcal{B}_2$ & $-0.01087479(1)$   & $2/15 \mathcal{B}_2$ & $-0.01087479(1)$   \\
$0.9$ & \multicolumn{2}{|c|}{$0.12894128(1) $} & $3/20 \mathcal{B}_2$ & $-0.01207857(1)$   & $3/20 \mathcal{B}_2$ & $-0.01207857(1)$   \\
$1.0$ & \multicolumn{2}{|c|}{$0.142461144(1)$} & $1/6  \mathcal{B}_2$ & $-0.013344189(1)$  & $1/6  \mathcal{B}_2$ & $-0.013344189(1)$  \\
$1.1$ & \multicolumn{2}{|c|}{$0.155819619(1)$} & $11/60\mathcal{B}_2$ & $-0.01465430(1) $  & $11/60\mathcal{B}_2$ & $-0.01465430(1)$   \\
$1.2$ & \multicolumn{2}{|c|}{$0.169016705(1)$} & $1/5  \mathcal{B}_2$ & $-0.01599680(1) $  & $1/5  \mathcal{B}_2$ & $-0.01599680(1)$   \\
$1.3$ & \multicolumn{2}{|c|}{$0.182052403(1)$} & $13/60\mathcal{B}_2$ & $-0.01736306(1) $  & $13/60\mathcal{B}_2$ & $-0.01736306(1)$   \\
$1.4$ & \multicolumn{2}{|c|}{$0.194926711(1)$} & $7/30 \mathcal{B}_2$ & $-0.018746916(1)$  & $7/30 \mathcal{B}_2$ & $-0.018746916(1)$  \\
$1.5$ & \multicolumn{2}{|c|}{$0.207639631(1)$} & $1/4  \mathcal{B}_2$ & $-0.02014387(1) $  & $1/4  \mathcal{B}_2$ & $-0.02014387(1)$   \\
\hline
%
\multicolumn{7}{|c|}{Brillouin/Sym.} \\
\hline
$r$ & \multicolumn{2}{|c|}{(a)}& \multicolumn{2}{|c|}{(b)}& \multicolumn{2}{|c|}{(c)} \\
\hline
$0.5$ & $-1/6  \mathcal{B}_2$ & $0.0049948(1)$  & $-9/4   \mathcal{B}_2$ & $0.0535829(1)$ & $9/4   \mathcal{B}_2$ & $-0.05491268(1)$ \\
$0.6$ & $-1/5  \mathcal{B}_2$ & $0.0061572(1)$  & $-27/10 \mathcal{B}_2$ & $0.063236(1)$  & $27/10 \mathcal{B}_2$ & $-0.06589522(1)$ \\
$0.7$ & $-7/30 \mathcal{B}_2$ & $0.0073273(1)$  & $-63/20 \mathcal{B}_2$ & $0.0730149(1)$ & $63/20 \mathcal{B}_2$ & $-0.07687776(1)$ \\
$0.8$ & $-4/15 \mathcal{B}_2$ & $0.0084985(1)$  & $-18/5  \mathcal{B}_2$ & $0.082888(1)$  & $18/5  \mathcal{B}_2$ & $-0.0878603(1)$  \\
$0.9$ & $-3/10 \mathcal{B}_2$ & $0.00966762(1)$ & $-81/20 \mathcal{B}_2$ & $0.092832(1)$  & $81/20 \mathcal{B}_2$ & $-0.0988428(1)$  \\
$1.0$ & $-1/3  \mathcal{B}_2$ & $0.0108335(1)$  & $-9/2   \mathcal{B}_2$ & $0.102829(1)$  & $9/2   \mathcal{B}_2$ & $-0.1098254(1)$  \\
$1.1$ & $-11/30\mathcal{B}_2$ & $0.0119957(1)$  & $-99/20 \mathcal{B}_2$ & $0.112865(1)$  & $99/20 \mathcal{B}_2$ & $-0.1208079(1)$  \\
$1.2$ & $-2/5  \mathcal{B}_2$ & $0.0131542(1)$  & $-27/5  \mathcal{B}_2$ & $0.122933(1)$  & $27/5  \mathcal{B}_2$ & $-0.1317904(1)$  \\
$1.3$ & $-13/30\mathcal{B}_2$ & $0.0143091(1)$  & $-117/20\mathcal{B}_2$ & $0.133025(1)$  & $117/20\mathcal{B}_2$ & $-0.14277298(1)$ \\
$1.4$ & $-7/15 \mathcal{B}_2$ & $0.0154608(1)$  & $-63/10 \mathcal{B}_2$ & $0.143136(1)$  & $63/10 \mathcal{B}_2$ & $-0.1537555(1)$  \\
$1.5$ & $-1/2  \mathcal{B}_2$ & $0.0166094(1)$  & $-27/4  \mathcal{B}_2$ & $0.153262(1)$  & $27/4  \mathcal{B}_2$ & $-0.1647381(1)$  \\
\hline
$r$ & \multicolumn{2}{|c|}{(d)}& \multicolumn{2}{|c|}{(e)}& \multicolumn{2}{|c|}{(f)} \\
\hline
$0.5$ & \multicolumn{2}{|c|}{$0.05771118(1)$} & $1/12 \mathcal{B}_2$ & $-0.008049(1)$  & $1/12 \mathcal{B}_2$ & $-0.008049(1)$  \\
$0.6$ & \multicolumn{2}{|c|}{$0.06885251(1)$} & $1/10 \mathcal{B}_2$ & $-0.0089611(1)$ & $1/10 \mathcal{B}_2$ & $-0.0089611(1)$ \\
$0.7$ & \multicolumn{2}{|c|}{$0.07986020(1)$} & $7/60 \mathcal{B}_2$ & $-0.0100132(1)$ & $7/60 \mathcal{B}_2$ & $-0.0100132(1)$ \\
$0.8$ & \multicolumn{2}{|c|}{$0.09073426(1)$} & $2/15 \mathcal{B}_2$ & $-0.0111642(1)$ & $2/15 \mathcal{B}_2$ & $-0.0111642(1)$ \\
$0.9$ & \multicolumn{2}{|c|}{$0.10147468(1)$} & $3/20 \mathcal{B}_2$ & $-0.012386(1)$  & $3/20 \mathcal{B}_2$ & $-0.012386(1)$  \\
$1.0$ & \multicolumn{2}{|c|}{$0.1120815(1)$}  & $1/6  \mathcal{B}_2$ & $-0.013659(1)$  & $1/6  \mathcal{B}_2$ & $-0.013659(1)$  \\
$1.1$ & \multicolumn{2}{|c|}{$0.12255461(1)$} & $11/60\mathcal{B}_2$ & $-0.0149691(1)$ & $11/60\mathcal{B}_2$ & $-0.0149691(1)$ \\
$1.2$ & \multicolumn{2}{|c|}{$0.13289412(1)$} & $1/5  \mathcal{B}_2$ & $-0.016307(1)$  & $1/5  \mathcal{B}_2$ & $-0.016307(1)$  \\
$1.3$ & \multicolumn{2}{|c|}{$0.14309999(1)$} & $13/60\mathcal{B}_2$ & $-0.017666(1)$  & $13/60\mathcal{B}_2$ & $-0.017666(1)$  \\
$1.4$ & \multicolumn{2}{|c|}{$0.15317223(1)$} & $7/30 \mathcal{B}_2$ & $-0.019040(1)$  & $7/30 \mathcal{B}_2$ & $-0.019040(1)$  \\
$1.5$ & \multicolumn{2}{|c|}{$0.1631108(1)$}  & $1/4  \mathcal{B}_2$ & $-0.020426(1)$  & $1/4  \mathcal{B}_2$ & $-0.020426(1)$  \\
\hline
\end{tabular}
\caption{Contributions to ${\csw^{(1)}}_\mathrm{Brillouin}$ for $N_c=3$ from each diagram as a function of $r$.}
\label{tab:csw_perdiagram_brillouin}
\end{table}

\clearpage
\pagestyle{plain}

\end{document}